\documentclass[twocolumn,ngerman,english,aps,prd,amssymb,amsmath,showpacs]{revtex4-1}
\usepackage[T1]{fontenc}
\usepackage[latin9]{inputenc}
\usepackage{verbatim}
\usepackage{textcomp}
\usepackage{amsmath}
\usepackage{graphicx}
\usepackage{amssymb}
\usepackage{esint}

\makeatletter

\usepackage{mathrsfs}\usepackage{amsthm}\@ifundefined{definecolor}
 {\usepackage[usenames]{color}}{}
\definecolor{gelb}{rgb}{0.7,0.5,0.2}

\usepackage{hyperref}
\hypersetup{
    colorlinks,%
    citecolor=blue,%
    filecolor=blue,%
    linkcolor=magenta,%
    urlcolor=blue
}
\usepackage[all]{hypcap}

\newcommand{\rmd}{{\rm d}}
\newcommand{\rmdd}{{\rm dd}}
\newcommand{\rmddd}{{\rm ddd}}

\newcommand{\bg}{{\mathbf{g}}}

\newcommand{\BE}{{\mathbb{E}}}

\newcommand{\CA}{{\cal A}}

\newcommand{\cD}{{\cal D}}
\newcommand{\CE}{{\cal E}}

\newcommand{\CI}{{\cal I}}

\newcommand{\CR}{{\cal R}}

\newcommand{\CQ}{{\cal Q}}
\newcommand{\CW}{{\cal W}}
\newcommand{\SP}{{\mathscr P}}

\newcommand{\average}[1]{\left\langle #1 \right\rangle_\cD}
\newcommand{\laverage}[1]{\left\langle #1 \right\rangle_{\cD_{\rm \bf i}}}
\newcommand{\baverage}[1]{\left\langle #1 \right\rangle_{\cD_R}}
\newcommand{\aaverage}[1]{\left\langle #1 \right\rangle_{\CI}}
\newcommand{\caverage}[1]{\left\langle #1 \right\rangle_{\CI}}

\newcommand{\ltbaverage}[1]{\left\langle #1 \right\rangle_{\rm LTB}}
\newcommand{\averageDN}[1]{\left\langle #1\right\rangle _{\cD_{0}}}
\newcommand{\initial}[1]{{#1_{\rm \bf i}}}
\newcommand{\now}[1]{{#1_{\rm \bf 0}}}
\newcommand{\inI}{{\rm I}}
\newcommand{\inII}{{\rm II}}
\newcommand{\inIII}{{\rm III}}
\def\MarqueFin#1{\leavevmode\unskip\nobreak\quad\hspace*{\fill}{#1}}
\newcommand{\Marque}[2][black]{\MarqueFin{{\color{#1} #2}}}\def\SymPreuv{\boldmath$\square$}%
\newcounter{propositioncount}\newtheorem{propos1}[propositioncount]{Proposition}\newtheorem{propos2}[propositioncount]{Proposition}%

\makeatother

\usepackage{babel}

\begin{document}

\title{Lagrangian theory of structure formation in relativistic cosmology II:\\
average properties of a generic evolution model}

\author{Thomas Buchert$^{1}$, Charly Nayet$^{1}$ and Alexander Wiegand$^{2,3,1}$\smallskip}

\affiliation{$^1$Universit\'e de Lyon, Observatoire de Lyon, 
Centre de Recherche Astrophysique de Lyon, CNRS UMR 5574: Universit\'e Lyon~1 and \'Ecole Normale Sup\'erieure de Lyon, \\
9 avenue Charles Andr\'e, F--69230 Saint--Genis--Laval, France}

\affiliation{$^{2}$Max--Planck--Institut f\"ur Gravitationsphysik, Albert--Einstein--Institut,
 Am M\"uhlenberg 1, D--14476 Potsdam, Germany}

\affiliation{$^{3}$Fakult{\"a}t f{\"u}r Physik, Universit{\"a}t Bielefeld,
  Universit{\"a}tsstra{\ss}e 25, D--33615 Bielefeld, Germany
  \medskip \\
  Emails: buchert@obs.univ--lyon.fr, charly.nayet@gmail.com and
  alexander.wiegand@aei.mpg.de}
  
%
%

\begin{abstract}
Kinematical and dynamical properties of a generic inhomogeneous cosmological model, spatially averaged
with respect to free--falling (generalized fundamental) observers, are investigated for the matter model `irrotational dust'.
Paraphrasing a previous Newtonian investigation, we present a relativistic generalization 
of a backreaction model based on volume--averaging the `Relativistic Zel'dovich Approximation'. In this model we investigate the effect of  `kinematical backreaction' 
on the evolution of cosmological parameters as they are defined in an averaged
inhomogeneous cosmology, and we show that the backreaction model interpolates between orthogonal symmetry properties by covering subcases of the plane--symmetric solution, the Lema\^\i tre--Tolman--Bondi solution and the Szekeres solution. 
We so obtain a powerful model that lays the foundations for quantitatively addressing curvature inhomogeneities as they would be interpreted as `Dark Energy' or `Dark Matter' in a quasi--Newtonian cosmology. The present model, having a limited architecture due to an assumed FLRW background, is nevertheless capable of replacing $1/4$ of the needed amount for `Dark Energy' on domains of $200$ Mpc in diameter for typical (one--sigma) fluctuations in a CDM initial power spectrum. However, the model is far from explaining `Dark Energy' on larger scales (spatially), where a $6$\% effect on $400$ Mpc domains is identified that can be traced back to an on average negative intrinsic curvature today. One drawback of the quantitative results presented is the fact that the epoch when backreaction is effective on large scales and leads to volume acceleration lies in the future. We discuss this issue in relation to the initial spectrum, the `Dark Matter' problem, the coincidence problem, and the fact that large--scale `Dark Energy' is an effect on the past light cone (not spatial),
and we pinpoint key elements of future research.
\end{abstract}
\maketitle

\section{Introduction}

In a previous work \cite{rza1} (henceforth cited as [L1]) we developed
a Lagrangian theory for the evolution of structure in irrotational dust continua in the framework of general relativity.
This theory is characterized by employing as a single dynamical variable the set of spatial
coframes $\boldsymbol{\eta}^a$, in the comoving--synchronous metric
form:
\begin{equation}
^{(4)}\bg = -{\bf d}t^2 + \, ^{(3)} \bg \;\;;\;\; ^{(3)} \bg = \delta_{ab} \,\boldsymbol{\eta}^a \otimes \boldsymbol{\eta}^b =  g_{ij}\, {\bf d}X^i \otimes {\bf d}X^j \, ,\nonumber
\end{equation}
with the spatial metric coefficients $g_{ij} = \delta_{ab} \,\eta^a_{\;\,i} \eta^b_{\;\,j}$,
where $X^i$ are Gaussian normal (Lagrangian) coordinates that are constant
along flow lines (here geodesics); $i,j,k = 1,2,3$ denote coordinate indices, while $a,b,c = 1,2,3$ are just counters.
Having written Einstein's equations in terms of the variable $\boldsymbol{\eta}^a$, we are entitled to set up a perturbation scheme that only perturbs this variable, while other variables like the density, the metric or the curvature, are expressed as functionals of this perturbation.
Such a point of view extrapolates these latter physical quantities beyond a strict perturbative expansion. In this spirit we defined and evaluated the first--order perturbation scheme on a FLRW background cosmology as a clear--cut definition of what has been known as the "Relativistic Zel'dovich Approximation'' (henceforth abbreviated by RZA) in previous work (see \cite{rza1} and references on previous work therein, in particular \cite{kasai95,matarrese&terranova,russ:rza,russ:age,RZAandLTB}).

In the present work we will model the fluctuations using this approximation and combine the resulting backreaction model with exact average properties of inhomogeneous dust cosmologies. (This framework is reviewed in \cite{buchert:darkenergy} and \cite{buchert:focus} and will be briefly recalled below.)  We will so paraphrase a
detailed investigation in Newtonian cosmology \cite{bks}
(henceforth cited as [BKS]), where we presented the results of an in--depth quantitative
study of the \textit{kinematical backreaction effect} in Newtonian
cosmology \cite{buchertehlers}.  In order to evaluate the influence of
inhomogeneities on effective cosmological parameters we employed in \cite{bks} exact
solutions with planar and spherical symmetry, as well as
perturbative solutions for generic initial conditions in the Eulerian
and Lagrangian first--order perturbation schemes. The most general,
nonperturbative model studied was based on the first--order Lagrangian deformation that
was used as input into the
kinematical backreaction functional (fluctuation model), but the effective cosmology (the average) was then
evaluated within the framework of the exact averaged equations. We will here provide the corresponding model in general relativity.

In \cite{bks} our initial data setting was conservative, i.e. we started with a
standard \textit{Cold Dark Matter} power spectrum without a
cosmological constant at the beginning of the matter--dominated era, maintaining the Early Universe description as in the standard model. 
Perhaps the most surprising outcome of this study was that,
although the kinematical backreaction term $\CQ_{\cD}$ itself was
quantitatively negligible on large expanding domains, the other
cosmological parameters could still experience significant changes,
e.g. for the deceleration parameter we found typical fluctuations up to $30$
percent on the scale of $200$~Mpc (with $h=0.5$).
In this work we will give the results for two background models: (i) a CDM spectrum with $h=0.5$ as
in \cite{bks} to allow for comparison, and (ii) a $\Lambda$CDM spectrum with $h=0.7$ to study the effect of the 
background. For the interpretation of the emerging curvature inhomogeneities as `Dark Energy' we refer to the
first model. Note that both homogeneous--isotropic models are assumed to have Euclidean space sections.
Recall that, in Newtonian cosmology, the
kinematical backreaction vanishes by construction on the scale of  an imposed torus
architecture (i.e. the use of periodic boundary conditions) \cite{buchertehlers}, a restriction that no longer applies to
the general relativistic model. The present investigation thus offers a more general view on the backreaction
model, notably by allowing for a nontrivial curvature evolution.

We may approach the present work with the following guidelines:
\begin{itemize}
\item The {\em cosmic triangle} \cite{bahcall:triangle} of the three
  parameters of the standard model (related to the homogeneous
  density, the constant curvature and the cosmological constant) {\em
    fluctuates} on a given spatial scale. The measure of deviations
  from the FLRW time evolution of these parameters is a kinematical
  backreaction functional $\CQ_{\cD}$ that encodes
  averaged fluctuations in kinematical invariants (the rate of
  expansion and the rate of shear).
\item The influence of backreaction may be twofold: for dominating
  shear fluctuations the effective density decelerating the expansion  is
  larger than the actual matter density, thus mimicking the effect of a {\em
    Kinematical Dark Matter}; for dominating expansion fluctuations
  backreaction acts accelerating, thus mimicking the effect of a
  (positive) cosmological constant, i.e. {\em Kinematical Dark
    Energy}. 
\item In general relativity, backreaction entails an emerging intrinsic curvature. Thus,  
as soon as we require, on some scale, that the average model is given by
a standard FLRW model geometry, then the emerging curvature term allows to interpret the above--mentioned kinematical
effects as `Dark Matter' or `Dark Energy', respectively.
    In Newtonian cosmology, the backreaction term should be interpreted as cosmic variance of 
    velocity fluctuations on a Euclidean space;
    a general--relativistic model entails deviations from a Euclidean model geometry;
    we so are led to interpret the dark sources in an assumed Euclidean background as "curvature energies"  of the actual average
    distribution. 
\item A genuinely relativistic property is the coupling of the
  kinematical backreaction functional to the averaged scalar (3--Ricci)
  curvature.  This coupling is absent in weakly perturbed FLRW
  spacetimes,  e.g.  in a Newtonian framework where a fluctuating
  scalar curvature is absent, or in a post--Newtonian framework
  and standard relativistic perturbation theory at first order, where the
  cosmological model is required to stay close to the
  constant--curvature FLRW spacetime, if this latter is taken as the
  background spacetime.
\item If kinematical backreaction is positive, i.e. dominated by
  expansion fluctuations and by a negative averaged curvature, a small source term is capable of driving
  the averaged system away from the homogeneous--isotropic FLRW
  background. We found, with the help of a detailed dynamical systems analysis of scaling solutions
  \cite{phasespace}, generalizing a previous analysis \cite{morphon}, that the standard model is a {\em
    repellor} in this situation, and the physical background (the average) is no longer given by the FLRW background. We may say that the 
    physical background lies in the `Dark Energy sector'.
    A similar situation occurs in the case of shear--dominated evolution of small--scale domains with positive averaged curvature, where the average is driven to
    lie in the `Dark Matter sector'.
    This insight suggests, that a perturbation theory has to be set up on a background that interacts with structure formation
    (a framework for such a perturbation theory has been suggested \cite{roy:perturbations}). This remark implies that the 
    present investigation is limited, since we employ perturbations on a FLRW background. The resulting backreaction model,
    however, can be formally employed also in such a more general framework, which is the subject of forthcoming work.
\end{itemize}
    
We proceed as follows.  
In Sect.~\ref{sect:average-theory} we briefly review the equations governing the 
averaged inhomogeneous dust
cosmologies in the relativistic framework and provide different
representations of the equations that are used in this article. We
then move in Sect.~\ref{sect:quantifying} to dynamical models that
allow estimating the backreaction terms: we provide backreaction and curvature models for the Lagrangian
linear perturbation theory, restricted to a relativistic form of
Zel'dovich`s approximation investigated in detail in \cite{rza1}. Sect.~\ref{sect:consistency} is 
devoted to consistency checks including the Newtonian limit. The
plane, spherical and quasispherical Szekeres solutions serve as exact reference models in Sect.~\ref{sect:exact}.
We then illustrate the results in
Sect.~\ref{sect:evolution-cosmo} in terms of the time evolution of
cosmological parameters: we examine the evolution of the volume scale
factor for typical initial conditions; then we explore the cosmological density parameters
in the averaged model. Here, we also quantify the emerging intrinsic curvature.
We discuss the results and conclude in Sect.~\ref{sect:discussion}.


\section{Averaged Dust Model in General Relativity}

\label{sect:average-theory}

\subsection{Averaging}

In this section we will briefly set up the averaging framework developed in
\cite{buchert:onaverage1,buchert:onaverage2,buchert:static}.

Given a foliation of spacetime into flow--orthogonal hypersurfaces, we
investigate fluid averaging for the matter model `irrotational dust';
the geometry of the dust continuum is described by the spatial metric coefficients $g_{ij}$ in
the comoving and synchronous line element: 
\begin{equation}
  \rmd s^{2}= - \rmd t^{2}+g_{ij} \,\rmd X^{i} \rmd X^{j}\;\;,
\end{equation}
where $X^{i}$ are the already introduced (Lagrangian) coordinates that are constant along flow lines
(spacetime geodesics of freely falling observers). Proper
time derivative along the $4-$velocity $u^{\mu}$, that is here equal to the coordinate time derivative, will be denoted by
$\partial_{t}:=u^{\mu}\partial_{\mu}$ and sometimes by an
overdot (where greek indices label spacetime and latin ones spatial
coordinates).

We look at the scalar parts of Einstein's equations, averaged over a
compact spatial domain $\cD$ at time $t$ that evolved out of the
initial domain $\initial{\cD}$ at time $\initial t$, conserving the
material rest mass inside the domain. The volume scale factor $a_{\cD}$, depending
on content, shape and position of the domain $\cD$, is defined via the
domain's volume $V_{\cD}(t)=|\cD|$, and the initial volume
$V_{\initial{\cD}}=V_{\cD}(\initial t)=|\initial{\cD}|$:
\begin{equation}
  a_{\cD}(t):=\left(\frac{V_{\cD}(t)}{V_{\initial{\cD}}}\right)^{1/3}\;\;,
\label{eq:def-ad}
\end{equation}%
where the volume of the domain is given by: 
\begin{equation}
  V_{\cD}(t):=\int_{\cD}\;\rmd\mu_{g}\;\;,
\end{equation}%
with the \emph{Riemannian} volume element $\rmd \mu_{g}=J \rmd^{3}X$, and
the local volume deformation $J:=\sqrt{\det(g_{ij})/\det(G_{ij})}$,
$G_{ij}=g_{ij}(\initial t)$.

  For domains $\cD$, which keep the total material mass $M_{\cD}$
  constant, as for Lagrangian domains with geodesic motion of their
  boundaries, the average density is simply given by: 
\begin{equation}
    \average{\varrho}=\frac{\laverage{\varrho(\initial
        t)}}{a_{\cD}^{3}}=\frac{M_{\cD}}{a_{\cD}^{3}V_{\initial{\cD}}}\;\;;\;\;
    M_{\cD}=M_{\initial{\cD}}\;\;.
\label{eq:average-rho}
\end{equation}
In this setting we can derive equations analogous to the averaged equations derived in Newtonian cosmology
\cite{buchertehlers} (restricting attention to irrotational dust
flows). Spatial averaging of a scalar field $\Psi$ is defined
by: 
\begin{equation}
  \average{\Psi(t,X^{k})}:=\frac{1}{V_{\cD}}\int_{\cD}\rmd\mu_{g}\;\;\Psi(t,X^{k})\;\;.
\label{eq:average-GR}
\end{equation}
The key property of inhomogeneity of the field $\Psi$ is revealed by
the {\em rule of noncommutativity} \cite{buchertehlers,buchert:onaverage1}:
\begin{equation}
  \partial_{t}\average{\Psi(t,X^{k})}-\average{\partial_{t}\Psi(t,X^{k})}=\average{\Theta\Psi}-\average{\Theta}\average{\Psi}\;,
\label{eq:commutation1}
\end{equation}
where $\Theta$ denotes the trace of the expansion tensor. In the present spacetime setting the latter
is, up to the sign, the extrinsic curvature tensor $\Theta_{ij}=-K_{ij}$.
$\Theta$ describes the local rate of volume change of fluid elements along
the vector field $\boldsymbol{\partial}_{t}$. 

It is formally convenient in the following calculations, but not necessary, to introduce the following formal average:
\begin{equation}
  \aaverage{\CA}=\frac{1}{V_{\initial\cD}}\int_{\cD} \,\rmd^{3}X\
  \CA\;,
\label{eq:average-chart}
\end{equation}
where $V_{\initial\cD}=\int_{\initial\cD}\rmd^{3}X = \int_{\initial\cD} J (X^i , \initial{t})\rmd^{3}X $ is the volume of the initial domain 
$\initial\cD$ for $J(X^i , \initial{t}) =1$ (we will explain
below that this latter property of the local volume deformation holds for non--normalized coframes). This average coincides with
the Riemannian volume average, if we consider fields at initial time, $\CA = \CA (\initial{t})$ (therefore we label it by $\CI$).
Rewriting the Riemannian average in terms of this formal average operation, we obtain the useful formula:
\begin{equation} \average{\CA}=\frac{\aaverage{{\cal
      A}J}}{\aaverage{ J}}\;\;;\;\; \aaverage{J} = a_{\cD}^{3} \;.
      \label{eq:average--domain-vs-card}
\end{equation}
It will simplify the following calculations and is useful to discuss the Newtonian limit (see
Section~\ref{sect:newtonlimit}). 

\subsection{Equations for the evolution of average characteristics}

\subsubsection{Generalized Friedmann equations}

The spatially averaged equations for the volume scale factor
$a_{\cD}$,
respecting rest mass conservation within the domain $\cD$, read \cite{buchert:onaverage1}:\\
{\em averaged Raychaudhuri equation}:
\begin{equation}
  3\frac{{\ddot{a}}_{\cD}}{a_{\cD}}+4\pi
  G\frac{M_{\initial{\cD}}}{V_{\initial{\cD}}a_{\cD}^{3}}-\Lambda=\CQ_{\cD}\;;
\label{eq:expansion-law-GR}
\end{equation}
  {\em averaged Hamilton
  constraint}: 
\begin{equation}
  \left(\frac{{\dot{a}}_{\cD}}{a_{\cD}}\right)^{2}-\frac{8\pi
    G}{3}\frac{M_{\initial{\cD}}}{V_{\initial{\cD}}a_{\cD}^{3}}+\frac{\average{\CR}}{6}-\frac{\Lambda}{3}=-\frac{\CQ_{\cD}}{6}\;,
\label{eq:hamiltonconstraint}
\end{equation}%
where the total rest mass $M_{\initial{\cD}}$, the averaged spatial 3--Ricci
scalar $\average{\CR}$ and the {\em kinematical backreaction} $\CQ_{\cD}$ are domain--depen\-dent and, except the mass,
time depen\-dent functions. The kinematical backreaction source term
itself is given by: 
\begin{equation} \CQ_{\cD}:=2\average{\inII}-\frac{2}{3}\average{\inI}^{2}=\frac{2}{3}\average{\left(\Theta-\average{\Theta}\right)^{2}}-2\average{\sigma^{2}}\;.
\label{eq:Q-GR}
\end{equation}
Here, $\inI$ and $\inII$ denote the principal scalar invariants (defined equivalently
to Eq.~(\ref{eq:DefInv}) below) of the extrinsic curvature coefficients $K_{ij}$.  The second equality follows by introducing the
decomposition of the extrinsic curvature into the kinematical
variables, by means of the identity $\Theta_{\;\, j}^{i}:=-K_{\;\, j}^{i}$. The kinematical variables are the {\it rate of expansion}
$\Theta:=-K_{\;\, k}^{k}$ and the {\it shear} $\sigma_{\;\, j}^{i}=-K_{\;\, j}^{i}-\frac{1}{3}\Theta\delta_{\;\, j}^{i}$.  We also defined the
{\it rate of shear} $\sigma^{2}:=1/2\;\sigma_{\;\, j}^{i}\sigma_{\;\, i}^{j}$.

\noindent We appreciate a close correspondence of the GR equations
(\ref{eq:expansion-law-GR}) and (\ref{eq:hamiltonconstraint}) with
their Newtonian counterparts (see \cite{bks}). The first equation
is formally identical to the Newtonian one, while the second delivers
an additional relation between the averaged scalar curvature and the
kinematical backreaction term that has no Newtonian analogue. This
implies an important difference that becomes manifest by looking at
the time derivative of Eq.~(\ref{eq:hamiltonconstraint}). The
integrability condition that this time derivative agrees with
Eq.~(\ref{eq:expansion-law-GR}) is nontrivial in the inhomogeneous GR context and
reads:
\begin{equation}
  \partial_{t}\CQ_{\cD}+6 H_{\cD} \CQ_{\cD}+\partial_{t}\average{\CR}+2 H_{\cD} \average{\CR}=0\;\;,
\label{eq:integrability-GR}
\end{equation}
where $H_{\cD}:={\dot{a}}_{\cD}/a_{\cD}$ denotes the {\em volume Hubble functional}. 

Eq.~(\ref{eq:integrability-GR}) shows that averaged scalar curvature
and the kinematical backreaction term are directly coupled, unlike
in the Newtonian case, where the curvature is nonexistent. For initially
vanishing $k$ in the standard FLRW cosmology, the scalar curvature
remains zero. This is not the case in the inhomogeneous GR context, where kinematical
backreaction produces averaged curvature in the course of structure formation,
even for domains that are on average flat initially. (Note also that the sign of the averaged curvature
may change during the evolution contrary to FLRW models.) To express the
deviation of cosmic curvature from a constant--curvature  (quasi--Friedmannian) behavior, we
define the average peculiar--scalar curvature by
\begin{equation}
  {\cal W}_{\cD}:=\average{\CR}-6k_{\initial{\cD}}/a_{\cD}^{2}\;.
\label{eq:avpeccurv}
\end{equation}
Integrating Eq.~(\ref{eq:integrability-GR}) with this definition inserted, we obtain:
\begin{equation}
  \frac{1}{3a_{\cD}^{2}}\left(\frac{c_{\initial{\cD}}}{2}-\intop_{\initial t}^{t}\,\rmd t'\;\CQ_{\cD}\;\frac{\rmd}{\rmd t'}a_{\cD}^{2}(t')\right)=\frac{1}{6}\left(\CW_\cD +\CQ_{\cD}\right)\ ,
\label{eq:integrability-integral-GR}
\end{equation}
\noindent with $c_{\initial{\cD}}=\CW_{\initial\cD}+\CQ_{\initial{\cD}}$.  Inserting it back into
Eq.~(\ref{eq:hamiltonconstraint}) we obtain the \textit{generalized
  Friedmann equation}: 
\begin{eqnarray}
  \frac{\dot{a}_{\cD}^{2}+k_{\initial{\cD}}}{a_{\cD}^{2}}-\frac{8\pi
    G\average{\varrho}}{3}-\frac{\Lambda}{3}=\nonumber\\
    \frac{1}{3a_{\cD}^{2}}
    \left(\frac{c_{\initial{\cD}}}{2}-\intop_{\initial t}^{t}\,\rmd t'\;\CQ_{\cD}\;\frac{\rmd}{\rmd t'}a_{\cD}^{2}(t')\right) .
\label{eq:averagefriedmann}
\end{eqnarray}
This equation is formally equivalent to its Newtonian counterpart
\cite{buchertehlers}. It shows that, by eliminating the averaged
curvature, the whole history of the averaged kinematical fluctuations
acts as a source of a generalized Friedmann equation.
  This equation was the starting point of our investigations in \cite{bks}. Since it is also valid in general relativity,
  we are in the position to translate many results from \cite{bks} into the GR context.
  In particular, our numerical codes can be accordingly applied.

  We now provide a compact form of the averaged equations
  introduced above, as well as some derived quantities that we will analyze in this paper.
 
  \subsubsection{Effective Friedmannian framework}

  The above equations can formally be recast into standard
  Friedmann equations for an effective perfect fluid
  energy momentum tensor with new effective sources
  \cite{buchert:onaverage2}:
\begin{eqnarray}
    \varrho_{{\rm eff}}^{\cD}=\average{\varrho}&-\frac{1}{16\pi G}\CQ_{\cD}-\frac{1}{16\pi G}\CW_\cD \;\;;\nonumber \\
    {p}_{{\rm eff}}^{\cD}=&-\frac{1}{16\pi G}{\cal
      Q}_{\cD}+\frac{1}{48\pi
      G}\CW_\cD \;\;.
\label{eq:equationofstate}
\end{eqnarray}
\begin{eqnarray}
    &3\frac{{\ddot{a}}_{\cD}}{a_{\cD}}=\Lambda-4\pi G(\varrho_{{\rm eff}}^{\cD}+3{p}_{{\rm eff}}^{\cD})\;\;;\nonumber\\
    &3H_{\cD}^{2}+ \frac{3 k_{\initial\cD}}{a_\cD^2}=\Lambda+8\pi G\varrho_{{\rm eff}}^{\cD}\;\;;\nonumber \\
    &{\dot{\varrho}}_{{\rm eff}}^{\cD}+3H_{\cD}\left(\varrho_{{\rm eff}}^{\cD}+{p}_{{\rm eff}}^{\cD}\right)=0\;\;.
\label{eq:effectivefriedmann}
\end{eqnarray}%
  Eqs.~(\ref{eq:effectivefriedmann}) correspond to Eqs.~(\ref{eq:expansion-law-GR}), (\ref{eq:hamiltonconstraint}) and (\ref{eq:integrability-GR}),
  respectively.

  This system of equations does not close unless we impose a model for
  the inhomogeneities. Note that, if the system would close, this
  would mean that we solved the scalar parts of the GR equations in
  general by reducing them to a set of ordinary differential equations
  on arbitrary scales $\cD$. Closure assumptions have been studied by
  prescribing a \textit{cosmic equation of state} of the form $p_{{\rm
      eff}}^{\cD}=\beta(\varrho_{{\rm eff}}^{\cD},a_{\cD})$
  \cite{buchert:darkenergy,buchert:static}, or by prescribing the
  backreaction terms through {\em scaling solutions}, e.g. ${\cal
    Q}_{\cD}\propto a_{\cD}^{n}$ \cite{morphon,phasespace}. In this paper we are
  going to explicitly model $\CQ_{\cD}$ by a relativistic
  Lagrangian perturbation scheme.

\subsubsection{The cosmic quartet and derived parameters}

For later convenience we introduce the set of dimensionless
average characteristics:
\begin{eqnarray}
  \Omega_{m}^{\cD}:=\frac{8\pi G}{3H_{\cD}^{2}}\average{\varrho}\;\;;\;\;\Omega_{\Lambda}^{\cD}:=\frac{\Lambda}{3H_{\cD}^{2}}\;\;;\nonumber \\
  \Omega_{\CR}^{\cD}:=-\frac{\average{\CR}}{6H_{\cD}^{2}}\;\;; \;\;\Omega_{\CQ}^{\cD}:=-\frac{\CQ_{\cD}}{6H_{\cD}^{2}}\;\;.
\label{eq:cospar}
\end{eqnarray}
We will, henceforth, call these characteristics
`parameters', but the reader should keep in mind that these are
functionals on $\cD$.  Expressed through these parameters the averaged
Hamilton constraint (\ref{eq:hamiltonconstraint}) assumes the form
of a \textit{cosmic quartet} \cite{buchert:onaverage}:
\begin{equation}
  \Omega_{m}^{\cD}\;+\;\Omega_{\Lambda}^{\cD}\;+\;\Omega_{\CR}^{\cD}\;+\;\Omega_{\CQ}^{\cD}\;=\;1\;\;.
\label{hamiltonomega}
\end{equation}
In this set, the averaged scalar curvature parameter and the kinematical
backreaction parameter are directly expressed through $\average{\CR}$ and $\CQ_{\cD}$, respectively.

  In order to compare these parameters with the `Friedmannian
  curvature parameter' that had to be used in Newtonian cosmology
  \cite{bks}, and that is employed to interpret observational data,
  we can alternatively introduce the set of parameters
\begin{eqnarray}
    \Omega_{k}^{\cD}:=-\frac{k_{\initial{\cD}}}{a_{\cD}^{2}H_{\cD}^{2}}= \Omega^\cD_\CR - \Omega^\cD_\CW \;\;;\;\; \Omega_\CW^\cD : = \frac{- \CW_\cD}{6 H_\cD^2}\;;\nonumber\\
    \Omega_{\CQ N}^{\cD}:=
    \left(\frac{c_{\initial{\cD}}}{2}-\intop_{\initial t}^{t}\,\rmd t'\;\CQ_{\cD}\;\frac{\rmd}{\rmd t'}a_{\cD}^{2}(t')\right)\,, \quad\quad   
  \label{omeganewton}
\end{eqnarray}
 being related to the previous parameters by
\begin{equation}
    \Omega_{k}^{\cD}+\Omega_{\CQ N}^{\cD}\;=\;\Omega_{\CR}^{\cD}+\Omega_{\CQ}^{\cD}\;\;.
\label{parameterrelation}
\end{equation}
Like the volume scale factor $a_{\cD}$ and the volume
Hubble functional $H_{\cD}$, we may introduce `parameters' for higher
derivatives of the volume scale factor, e.g. the {\em volume deceleration functional}
\begin{equation}
  q^{\cD}:=-\frac{{\ddot{a}}_{\cD}}{a_{\cD}}\frac{1}{H_{\cD}^{2}}=\frac{1}{2}\Omega_{m}^{\cD}+2\Omega_{\CQ}^{\cD}-\Omega_{\Lambda}^{\cD}\;\;.%
\label{deceleration}
\end{equation}
(For higher derivatives such as the {\em state finders}, see \cite{buchert:darkenergy}.)
In this paper we denote all the parameters evaluated at the initial
time by the index $\initial{\cD}$, and at the present time by the
index $\now\cD$.

\section{Quantifying Backreaction}

\label{sect:quantifying}

In the last section we saw how the appearance of the backreaction term
modifies the standard Friedmannian evolution. In the following we will
quantify this departure. Exact inhomogeneous solutions for estimating
the amount of backreaction are only available for highly symmetric
models like for models with plane symmetry
(Subsect.~\ref{sect:plane-symmetry}) and the spherically--symmetric
solutions (Subsect.~\ref{sect:spherical-symmetry}).  In generic
situations we have to rely on approximations. As the Newtonian result
\cite{bks} suggests, we will obtain also here an approximation
that interpolates in special cases between the above two subclasses of
exact GR solutions with orthogonal kinematical properties. However,
the classes of solutions will be more strongly restricted than in the
Newtonian case, because of the Hamilton constraint that joins the
system of equations in the GR context.

\subsection{Modeling the local deformation}

\subsubsection{Representation of deformations through coframes}

To be able to use the `Relativistic Zel'dovich Approximation'
as formulated in \cite{rza1}, we will first express
the kinematical backreaction term of Eq.~(\ref{eq:Q-GR}) in terms of
the spatial coframe coefficients $\eta_{\; i}^{a}\;$.  These latter are defined through the
metric form coefficients
\begin{equation} g_{ij}\left(t,X^{k}\right):=G_{ab}\,\eta_{\;
    i}^{a}\left(t,X^{k}\right)\eta_{\;
    j}^{b}\left(t,X^{k}\right)\;,
\end{equation}
where
$G_{ab}\,\delta_{\; i}^{a}\delta_{\; j}^{b}=G_{ij}=g_{ij}\left(\initial
  t,X^{k}\right)$ are the initial metric coefficients.  The coframes therefore contain the complete time evolution
of the metric. We now express the expansion
tensor, \[ \Theta_{\; j}^{i}=g^{ik}\dot{g}_{kj}\;,\] with the help of the
coframes. Since the inverse metric is analogously decomposable
into frames $e_{a}^{\; i}$ (that are inverse to the coframes $e_{a}^{\; i}\eta_{\; j}^{a}=
\delta_{\; j}^{i}\;\;;\;\; e_{a}^{\; i}\eta_{\; i}^{b}= \delta_{a}^{\; b}$), 
\begin{equation}
 g^{ij}=G^{ab}\,e_{a}^{\; i}e_{b}^{\; j}\;,
 \end{equation}
we can finally write (for details see \cite{rza1}):
\begin{equation} \Theta_{\; j}^{i}=e_{a}^{\; i}{\dot{\eta}}_{\;
    j}^{a}=\frac{1}{2J}\;\epsilon_{abc}\epsilon^{ik\ell}\,{\dot{\eta}}_{\;
    j}^{a}\,\eta_{\;
    k}^{b}\,\eta_{\;\ell}^{c}\;,
\label{expansionmatrix}
\end{equation}
with the local volume deformation (corresponding to the Jacobian in
Newtonian theory)
\begin{equation} J:=\det(\eta_{\;
    i}^{a})=\frac{1}{6}\epsilon_{abc}\epsilon^{ijk}\,\eta_{\;
    i}^{a}\eta_{\; j}^{b}\eta_{\;
    k}^{c}\;.%
\label{jacobian}
\end{equation}
This may then be used to
calculate the first two invariants of $\Theta_{\; j}^{i}$ and to obtain
the expression
\begin{equation}
\CQ_{\mathcal{D}}=\frac{1}{\aaverage J}\aaverage{\epsilon_{abc}\epsilon^{ikl} \dot{\eta}_{\;i}^{a} \dot{\eta}_{\;k}^{b} \eta_{\;l}^{c}}-\frac{2}{3}\left(\frac{\aaverage{\dot{J}}}{\aaverage J}\right)^{2},
\end{equation}
for $\CQ_{\cD}$ that solely depends on the coframes
$\eta_{\; i}^{a}$ and their time derivatives.

\subsubsection{The `Relativistic Zel'dovich Approximation'}
\label{sec:RZA}
\label{sect:backreactionkin} The equations and `parameters' introduced
in Section \ref{sect:average-theory} can live without introducing a
background spacetime. The description is background--free and
nonperturbative.  Also expressing $\CQ_{\cD}$ in terms of coframes
as sketched in the previous section is still completely
general. However, in order to provide a concrete model for the
backreaction terms, we employ methods of perturbation
theory. We will, however, only model the fluctuations by perturbation
theory; by using the exact expressions for the
functionals in terms of coframes, we extrapolate the first--order perturbative solution.

We now introduce perturbed coframes
analogous to \cite{rza1}.
In perturbation schemes one usually defines a reference frame through
a known solution, e.g. a frame {\em comoving} with the Hubble flow
(i.e. a FLRW solution) that we now represent through three
homogeneous--isotropic deformation one--forms (labeled by $a,b,c...$
and expressed in the local exact coordinate basis labeled by
$i,j,k...$), 
\begin{equation} \boldsymbol{\eta}_{H}^{\;\; a}=\eta_{H\,
    i}^{\;\; a}\,\boldsymbol{\rmd}X^{i}:=a(t)\boldsymbol{\eta}_{H}^{\;\;
    a}(\initial t)\;,\;\;\;\;\eta_{H\, i}^{\;\; a}:=a(t)\delta^{\;
    a}_{\;\; i}\;\;,%
\label{hubbleform}
\end{equation}
 where $a(t)$ is
a solution of Friedmann's differential equation
\begin{equation}
    H^{2}=\frac{8\pi G\varrho_{H}(t)+\Lambda}{3}-\frac{k}{a^{2}(t)}\;,
\label{eq:fried-expansion}
\end{equation}
with $H(t):= \dot a / a$, $\varrho_{H}(t)=\varrho_{H_{\rm i}}/a^{3}(t)$ and $a(\initial t)=1$.
For the full deformation one--forms we prescribe the superposition 
\begin{equation}
  \boldsymbol{\eta}^{\, a}=\boldsymbol{\eta}_{H}^{\;\; a}+a(t){\bf
    P}^{\, a}\;,
\label{ansatz}
\end{equation}
with inhomogeneous
deformation one--forms ${\bf P}^{\, a}(t,X^{k})$.  To first order,
they can be restricted to the relativistic generalization of
Zel'dovich's approximation (RZA) \cite{rza1}:
\begin{equation} {\bf
    P}^{\, a}=\xi(t)\dot{P}_{\;
    i}^{a}\;\boldsymbol{\rmd}X^{i}\;\;,
\end{equation}
with $\dot{P}_{\;  i}^{a}:=\dot{P}_{\; i}^{a}\left(\initial t,X^{k}\right)$.  
The function $\xi\left(t\right)$ is defined by
\begin{equation}
  \xi(t):=\lbrack q(t)-q(\initial t)\rbrack/{\dot{q}(\initial t)}\ ,
\label{eq:def-xi}
\end{equation}
where the function $q(t)$ is solution of the equation:
\begin{equation}
  \ddot{q}(t)+2\frac{\dot{a}(t)}{a(t)}\dot{q}(t)+\left(3\frac{\ddot{a}(t)}{a(t)}-\Lambda\right)q(t)=0\,.
\label{eq:evolution-q}
\end{equation}
Thus, the function $\xi$ satisfies:
\begin{equation}
    \ddot{\xi}(t)+2\frac{\dot{a}(t)}{a(t)}\dot{\xi}(t)+\left(3\frac{\ddot{a}(t)}{a(t)}-\Lambda\right)\left(\xi(t)+\frac{q(\initial
        t)}{\dot{q}(\initial t)}\right)=0 \ .
\label{eq:evolution-xi}
\end{equation}
Explicit solutions of this equation for different
backgrounds may be found in \cite{buchert:class}, those including a
cosmological constant in \cite{bildhauer:solutions}.

Writing the RZA of Eq. (\ref{ansatz}) in the exact coordinate basis
${\bf d} X^{i}$ yields: 
\begin{equation} ^{{\rm RZA}}\eta_{\
    i}^{a}\left(t,X^{k}\right):=a\left(t\right)\left(N{^{a}}_{i}+\xi(t)\dot{P}_{\
      i}^{a}\right),
\label{eq:zeldovich-def}
\end{equation}
where
$N{^{a}}_{i}:={^{{\rm RZA}}\eta}_{\ i}^{a}\left(\initial
  t,X^{k}\right)$ and $\dot{P}_{\ i}^{a}=\dot{P}_{\
  i}^{a}\left(\initial t,X^{k}\right)$.

Before we use this expression to determine the backreaction term, let
us elaborate on a subtlety with the choice of initial conditions.
The expression of the metric tensor in terms of nonintegrable
coframes:
\begin{equation}
  g_{ij}:=G_{ab}\,\eta_{\; i}^{a}\eta_{\; j}^{b}\;,
\end{equation}
allows two different treatments of the initial displacements. One
can either include them into $G_{ab}$ which means that the noncoordinate
basis is orthogonal but not orthonormal. Or one can choose $G_{ab}$
to be orthonormal (being the standard assumption), i.e. $G_{ab} = \delta_{ab}$, but then one has to deal with the initial values of
the coframes. To have both at a time, i.e. $N{^{a}}_{i}=\delta_{\; i}^{a}$
and $G_{ab}=\delta_{ab}$, is not possible as this would mean that the
RZA initial metric would be Euclidean, and this would disable any
time evolution of this metric as pointed out in \cite{matarrese&terranova}
and \cite{russ:rza}. As a nontrivial time evolution of the metric
is what we are interested in, there are only two options:
\begin{itemize}
\item [{\it O}1:] If $\tilde{N}{^{a}}_{i}:=\delta_{\;
    i}^{a}+P{^{a}}_{i}$, with $P{^{a}}_{i}:=P_{\ i}^{a}(\initial
  t,X^{k})$ and $P{^{a}}_{i}\neq0$, then $\textbf{G}$ can be
  restricted to $\boldsymbol{\delta}$, and the coframes
  read $^{{\rm RZA}}\tilde{\eta}_{\
    i}^{a}\left(t,X^{k}\right):=a\left(t\right)\left(\delta_{\;
      i}^{a}+P{^{a}}_{i}+\xi(t)\dot{P}_{\ i}^{a}\right)$ with the metric $g_{ij}:=\delta_{ab}\,\tilde{\eta}_{\; i}^{a}\tilde{\eta}_{\; j}^{b}$.
\item [{\it O}2:] By appropriate coordinate transformations, one may
  set $N{^{a}}_{i}$ in Eq.~(\ref{eq:zeldovich-def}) to $\delta_{\;
    i}^{a}$; the transformation then sends $\dot{P}_{\
    i}^{a}\rightarrow\dot{\mathscr P}_{\ i}^{a}=\delta_{\;
    j}^{a}{}^{{\rm RZA}}\tilde{e}_{\ b}^{j}\left(\initial
    t,X^{k}\right)\dot{P}_{\ i}^{b}$, and all information about the
  initial geometrical inhomogeneities is contained in $\textbf{G}$.
  The coframes become $^{{\rm RZA}}\eta_{\
    i}^{a}\left(t,X^{k}\right):=a\left(t\right)\left(\delta_{\;
      i}^{a}+\xi(t)\dot{\mathscr P}_{\ i}^{a}\right)$, and the metric
  $g_{ij}:=G_{ab}\,\eta_{\; i}^{a}\eta_{\; j}^{b}$ with
  $G_{ab}=\delta_{cd}\,\tilde{N}{^{c}}_{a}\tilde{N}{^{d}}_{b}$, with $\tilde{N}^c_{\;a}$ as defined above.
\end{itemize}
In order to have a complete formal correspondence with the Newtonian model, we stick
to the second option in what follows. (Notice that we have chosen the first option in \cite{rza1}.)

\subsection{Kinematical backreaction and intrinsic curvature models}

After the definition of the perturbative setup explained in the previous
section, we now turn to the concrete calculation of the
backreaction and curvature models. As discussed in the previous section,
we use non--normalized coframes (Option 2; the corresponding expressions with Option 1 will be listed in appendix \ref{sec:Appendix-B}),
\begin{equation}
  g_{ij}=G_{ab}\,\eta_{\ i}^{a}\eta_{\ j}^{b}\ .
\end{equation}
As pointed out by Chandrasekhar \cite{chandra:blackholes} such a choice
can lead to formally simpler expressions in some cases, and here we
encounter such a case. Indeed, the expression for the kinematical
backreaction term turns out to resemble more closely its Newtonian
counterpart, and the formulas become more concise.

The `Relativistic Zel'dovich Approximation' is then defined as:
\begin{equation}
 ^{{\rm RZA}}\eta_{\ i}^{a}(t,X^{k}):=a(t)\left(\delta_{\ i}^{a}+\xi(t)\dot{\mathscr P}_{\ i}^{a}\right)\ ,\nonumber
\label{eq:etaRZA}
\end{equation}
 with: \begin{align}
\left\{ \begin{array}{lll}
    \dot{\mathscr P}_{\ i}^{a}=\dot{\mathscr P}_{\ i}^{a}(\initial t,X^{k})\ ,\\
    \xi(\initial t)=0\ ,\\
    a(\initial t)=1\ .\end{array}\right.
\end{align} 
By definition (see \cite{rza1}), the RZA of
any field is the evaluation of this field in terms of its functional
dependence on the RZA coframes, Eq.~(\ref{eq:etaRZA}), without
any further approximation or truncation. Therefore, the metrical distances are calculated exactly for the approximated deformation:
\begin{eqnarray}
  ^{{\rm RZA}}g_{ij}(t,X^{k}) & = & a^{2}(t)\left\{ G_{ij}+\xi(t)\left(G_{aj}\dot{\mathscr P}_{\ i}^{a}+G_{ib}\dot{\mathscr P}_{\ j}^{b}\right)\right.\nonumber \\
  & & \left.+\xi^{2}(t)G_{ab}\dot{\mathscr P}_{\ i}^{a}\dot{\mathscr
      P}_{\ j}^{b}\right\} \ ,
\end{eqnarray}
 and so
\begin{equation}
  ^{{\rm RZA}}g_{ij}(\initial t)=:G_{ij}\ .
\end{equation}
The local volume deformation, Eq.~(\ref{jacobian}), then becomes:
\begin{equation}
  ^{{\rm RZA}}{\rm J}=: a^{3}(t){\frak J}\ ,
\label{eq:RZAJac}
\end{equation}
(implying that for the approximated deformation we exactly conserve mass, which would not be the case in a strictly linearized setting);
we have introduced the \textit{peculiar--volume deformation},
following from (\ref{eq:etaRZA}),
\begin{equation}
  {\frak J}(t,X^{k}):=1+\xi(t){\rm I}_{{\rm {\bf i}}}+\xi^{2}(t){\rm II}_{{\rm {\bf i}}}+\xi^{3}(t){\rm III}_{{\rm {\bf i}}}\ ,
\label{eq:pecvolume}
\end{equation}
\begin{equation} {\rm I}_{{\rm {\bf i}}}:={\rm I}\left(\dot{\mathscr
      P}_{\ i}^{a}\right);\;{\rm II}_{{\rm {\bf i}}}:={\rm
    II}\left(\dot{\mathscr P}_{\ i}^{a}\right);\;{\rm III}_{{\rm {\bf
        i}}}:={\rm III}\left(\dot{\mathscr P}_{\
      i}^{a}\right)\;,
\label{eq:inInv}
\end{equation}
where the principal scalar invariants
of the matrix $\dot{\mathscr P}_{\ i}^{a}$ are given by
\begin{eqnarray}
  {\rm I}\left(\dot{\mathscr P}_{\ i}^{a}\right) & := & \frac{1}{2}\epsilon_{abc}\epsilon^{ijk}\dot{\mathscr P}_{\ i}^{a}\delta_{\ j}^{b}\delta_{\ k}^{c}\ ,\nonumber \\
  {\rm II}\left(\dot{\mathscr P}_{\ i}^{a}\right) & := & \frac{1}{2}\epsilon_{abc}\epsilon^{ijk}\dot{\mathscr P}_{\ i}^{a}\dot{\mathscr P}_{\ j}^{b}\delta_{\ k}^{c}\ ,\nonumber \\
  {\rm III}\left(\dot{\mathscr P}_{\ i}^{a}\right) & := &
  \frac{1}{6}\epsilon_{abc}\epsilon^{ijk}\dot{\mathscr P}_{\
    i}^{a}\dot{\mathscr P}_{\ j}^{b}\dot{\mathscr P}_{\ k}^{c}\ .
\label{eq:DefInv}
\end{eqnarray}

\subsubsection{The kinematical backreaction}

In addition to the local volume deformation $J$, we need the
invariants of the expansion tensor $\Theta_{\; j}^{i}$. As we have
introduced a background by Eq. (\ref{ansatz}), we can define a {\em
  peculiar--expansion tensor} $\theta_{\; j}^{i}:=\Theta_{\;
  j}^{i}-H(t)\delta_{\; j}^{i}$ with respect to this background. The
three principal scalar invariants of the full expansion tensor
decompose into invariants of the peculiar--expansion tensor
as follows:
\begin{eqnarray}
  & \inI(\Theta_{\; j}^{i})=3H+\inI(\theta_{\; j}^{i})\;,\nonumber \\
  & \inII(\Theta_{\; j}^{i})=3H^{2}+2H\;\inI(\theta_{\; j}^{i})+\inII(\theta_{\; j}^{i})\;,\nonumber \\
  & \inIII(\Theta_{\; j}^{i})=H^{3}+H^{2}\inI(\theta_{\
    j}^{i})+H\inII(\theta_{\ j}^{i})+\inIII(\theta_{\
    j}^{i})\;.
\label{eq:inIII-pelicular_inIII}
\end{eqnarray}
Inserting Eqs.~(\ref{eq:inIII-pelicular_inIII}) into Eq.~(\ref{eq:Q-GR}), we can
write the backreaction variable in terms of invariants of the peculiar--expansion tensor: 
\begin{equation}
\CQ_{\cD}=2\average{\inII(\theta_{\;
      j}^{i})}-\frac{2}{3}\average{\inI(\theta_{\;
      j}^{i})}^{2}.
\label{eq:backreaction-pelicular_tensor}
\end{equation}
This nontrivial result demonstrates that the backreaction effects do
not depend on an assumed homogeneous reference background (a Hubble flow): backreaction is only due to
inhomogeneities.  Now, using the formula,
Eq.~(\ref{eq:average--domain-vs-card}), the backreaction term
(\ref{eq:backreaction-pelicular_tensor}) can be expressed in terms of
the formal average (\ref{eq:average-chart}): 
\begin{equation} 
\CQ_{\cD}=\frac{2}{\aaverage{\frak J }^{2}}\left[\aaverage{{\rm II}(\theta_{\;
      j}^{i}){\frak J}}\aaverage{\frak J}-\frac{1}{3}\aaverage{{\rm I}(\theta_{\;
      j}^{i}){\frak J} }^{2}\right].
\label{eq:backreaction--card}
\end{equation}
Now we have all we need. Plugging Eq.~(\ref{eq:etaRZA})
into Eq.~(\ref{expansionmatrix}) and subtracting $H(t)\delta_{\;
  j}^{i}$, we obtain for the scalar invariants of the
peculiar--expansion tensor, 
\begin{equation} 
^{{\rm
        RZA}}{\rm I}(\theta_{\ j}^{i})=\frac{\dot{{\frak J}}}{{\frak J}}\ ,\
    \ ^{{\rm RZA}}{\rm II}(\theta_{\
      j}^{i})=\frac{1}{2}\left(\frac{\ddot{{\frak J}}}{{\frak J}}-\frac{\ddot{\xi}(t)}{\dot{\xi}(t)}\frac{\dot{{\frak J}}}{{\frak J}}\right)\
    ,\nonumber
\label{eq:RZA-peculiar-Inv}
\end{equation}
\begin{equation} 
^{{\rm RZA}}{\rm III}(\theta_{\
      j}^{i})=\frac{1}{6}\left(\frac{\dddot{{\frak J}}}{{\frak J}}-\frac{\dot{{\frak J}}}{{\frak J}}\frac{\dddot{\xi}(t)}{\dot{\xi}(t)}\right)-\frac{1}{2}\left(\frac{\ddot{{\frak J}}}{{\frak J}}-\frac{\dot{{\frak J}}}{{\frak J}}\frac{\ddot{\xi}(t)}{\dot{\xi}(t)}\right)\frac{\ddot{\xi}(t)}{\dot{\xi}(t)}\
    ,
\end{equation}
which implies for the backreaction term
\begin{equation} 
^{{\rm RZA}}\CQ_{\cD}=\frac{\aaverage{\ddot{{\frak J}}}}{\aaverage{{\frak J}}}-\frac{\ddot{\xi}}{\dot{\xi}}
\frac{\aaverage{\dot{{\frak J}}}}{\aaverage{{\frak J}}}-\frac{2}{3}\left(\frac{\aaverage{\dot{{\frak J}}}}{\aaverage{{\frak J}}}\right)^{2},
\label{eq:backreaction-RZA-1}
\end{equation}
which with Eq.~(\ref{eq:pecvolume}) yields a compact form of the RZA backreaction model: 
\begin{eqnarray}
  ^{{\rm RZA}}\CQ_{\cD}\;= & {\displaystyle
    \frac{\dot{\xi}^{2}\left(\gamma_{1}+\xi\gamma_{2}+\xi^{2}\gamma_{3}\right)}{\left(1+\xi\caverage{{\rm
          I}_{{\rm {\bf i}}}}+\xi^{2}\caverage{{\rm
          II}_{{\rm {\bf i}}}}+\xi^{3}\caverage{{\rm
          III}_{{\rm {\bf i}}}}\right)^{2}}\;,}\nonumber
\label{resultQ2}
\end{eqnarray}
with:
\begin{align}
  \begin{cases}
    \gamma_{1}:=2\caverage{{\rm II}_{{\rm {\bf i}}}}-\frac{2}{3}\caverage{{\rm I}_{{\rm {\bf i}}}}^{2},\\
    \gamma_{2}:=6\caverage{{\rm III}_{{\rm {\bf i}}}}-\frac{2}{3}\caverage{{\rm II}_{{\rm {\bf i}}}}\caverage{{\rm I}_{{\rm {\bf i}}}},\\
    \gamma_{3}:=2\caverage{{\rm I}_{{\rm {\bf i}}}}\caverage{{\rm III}_{{\rm {\bf i}}}}-\frac{2}{3}\caverage{{\rm II}_{{\rm {\bf
          i}}}}^{2}.\end{cases}
\label{eq:GammaDef}
\end{align}
We note that the initial metric tensor ${\bf G}$ does now not explicitly appear in the expression for
the backreaction model; the above expression is formally equivalent with the Newtonian expression \cite{bks}.

\subsubsection{The intrinsic curvature model}

Unlike in the Newtonian case \cite{bks}, where there exists a flat embedding space,
RZA is fully intrinsic (i.e. the structures are not embedded into an external space, but propagate within the space given by the coframe deformation), and we have a 
nonvanishing scalar $3-$curvature. Its expression may be found by combining the Hamilton constraint and the
Raychaudhuri equation: 
\begin{equation} 
\CR=6{\rm II}(\Theta_{\ j}^{i})-4{\rm I}^{2}(\Theta_{\ j}^{i})-4{\rm
    \dot{I}}(\Theta_{\ j}^{i})+6\Lambda\ ,
\end{equation}
or in terms of the first two scalar invariants of the peculiar--expansion
tensor: 
\begin{equation} 
\CR=6{\rm II}(\theta_{j}^{i})-4{\rm
    I}^{2}(\theta_{j}^{i})-4\dot{{\rm I}}(\theta_{j}^{i})-12H{\rm
    I}(\theta_{j}^{i})+\frac{6k}{a^{2}(t)}\;.
\label{eq:curvature-vs-peculiar-inv}
\end{equation}
For the averaged peculiar--scalar--curvature, Eq.~(\ref{eq:avpeccurv}), using
Eq.~(\ref{eq:RZA-peculiar-Inv}) and
Eq.~(\ref{eq:curvature-vs-peculiar-inv}), we get:
\begin{eqnarray}
  ^{{\rm RZA}}{\cal W}_{\cD}=-\left\{ \frac{\caverage{\ddot{{\frak J}}}}{\caverage{{\frak J}}}+3\left(\frac{\ddot{\xi}}{\dot{\xi}}+4\frac{\dot{a}}{a}\right)\frac{\caverage{\dot{{\frak J}}}}{\caverage{{\frak J}}}\right\} \nonumber \\
  +6\left(\frac{k}{a^{2}}-\frac{k_{\initial{\cD}}}{a_{\cD}^{2}}\right) .& 
\label{eq:RZA-deviation-term}
\end{eqnarray}
Inserting the result for ${\frak J}$ from Eq.~(\ref{eq:pecvolume}),
we find the explicit time dependence in a form similar to  $\CQ_{\cD}$ in Eq.~(\ref{resultQ2}):
\begin{eqnarray}
  ^{{\rm RZA}}{\cal W}_{\cD}\; & = & \frac{\dot{\xi}^{2}\left(\tilde{\gamma}_{1}+\xi\tilde{\gamma}_{2}+\xi^{2}\tilde{\gamma}_{3}\right)}{1+\xi\caverage{{\rm I}_{{\rm {\bf i}}}}+\xi^{2}\caverage{{\rm II}_{{\rm {\bf i}}}}+\xi^{3}\caverage{{\rm III}_{{\rm {\bf i}}}}}\nonumber \\
  &  & {\displaystyle +6\left(\frac{k}{a^{2}}-\frac{k_{\initial{\cD}}}{a_{\cD}^{2}}\right)\;,}\nonumber
\label{resultW}
\end{eqnarray}
with:\begin{align}
  \begin{cases}
    \tilde{\gamma}_{1}:=-2\caverage{\initial{\inII}}-12\caverage{\initial{\inI}}\frac{H}{\dot{\xi}}-4\caverage{\initial{\inI}}\frac{\ddot{\xi}}{\dot{\xi}^{2}}\,,\\
    \tilde{\gamma}_{2}:=-6\caverage{\initial{\inIII}}-24\caverage{\initial{\inII}}\frac{H}{\dot{\xi}}-8\caverage{\initial{\inII}}\frac{\ddot{\xi}}{\dot{\xi}^{2}}\,,\\
    \tilde{\gamma}_{3}:=-36\caverage{\initial{\inIII}}\frac{H}{\dot{\xi}}-12\caverage{\initial{\inIII}}\frac{\ddot{\xi}}{\dot{\xi}^{2}}\,.\end{cases}\end{align}

In this expression there are two important formal differences compared with the
functional form
of $\CQ_{\cD}$. First of all, the time derivatives of ${\xi}$ no
longer disappear automatically from the solution. In addition, the
curvature term also explicitly depends on the background via $H$. This
leads to a time dependence of the coefficients $\tilde{\gamma}_{i}$,
whereas for the kinematical backreaction functional they were simply constants. 

In the case of the Einstein--de Sitter (EdS) background, the growth of the first--order perturbation goes with the scale factor $q=a$, and so we can simplify the expression to
\begin{eqnarray}
^{{\rm RZA}}{\cal W}_{\cD}\; & = &\frac{-10\initial H\dot{\xi}^{2}}{a}\frac{\caverage{\initial{\inI}}+2\caverage{\initial{\inII}}\xi+3\caverage{\initial{\inIII}}\xi^{2}}{1+\xi\caverage{\initial{\inI}}+\xi^{2}\caverage{\initial{\inII}}+\xi^{3}\caverage{\initial{\inIII}}}\nonumber\\
&&-\dot{\xi}^{2}\frac{2\caverage{\initial{\inII}}+6\caverage{\initial{\inIII}}\xi}{1+\xi\caverage{\initial{\inI}}+\xi^{2}\caverage{\initial{\inII}}+\xi^{3}\caverage{\initial{\inIII}}} .
\end{eqnarray}
As $\dot{\xi}^2\propto a^{-1}$ in the EdS case, we recover the well--known result that the leading curvature contribution goes as $\propto a^{-2}$. The second term, which is the leading second order contribution, goes as $a^{-1}$ as expected \cite{li:scale}.
We will learn below that this form of the peculiar--curvature functional is not the best choice. As it
will turn out in the explicit consideration within the class of exact spherically symmetric solutions, Section~\ref{sect:LTB}, there exists a better approximation for 
numerical evaluations that employ $\CQ_{\cD}$ together with
the exact integrability constraint. This expression and the comparison with the above expression will be explicitly provided in Section~\ref{sect:curv}.

\section{Consistency checks}
\label{sect:consistency}

In order to evaluate the goodness of the averaged model based on the RZA deformation, we will
perform three consistency checks in this section.

\subsection{Newtonian limit}
\label{sect:newtonlimit}

First we show that the RZA result of Eq.~(\ref{resultQ2}) for the backreaction model
has the correct Newtonian limit. The procedure to
arrive at the Newtonian limit of the relativistic quantities follows the
prescription of \cite{rza1}, that sends the nonintegrable Cartan
coframes to integrable ones: \[ \eta_{\,\,
  i}^{a}\rightarrow\,^{N}\eta_{\,\,
  i}^{a}=f_{\,\,\left|i\right.}^{a}\;.\] As was demonstrated in \cite{rza1}, this
leads exactly to the Newtonian system of fluid equations in the
Lagrangian description. $f^{a}$ is the position vector field that maps
the Lagrangian coordinates to Eulerian positions, and its Lagrangian gradient
encodes the volume deformation of fluid elements.  In \cite{bks}
this has been approximated by the Newtonian form of the Zel'dovich approximation.
Comparing the corresponding expressions we find that \[ \,^{N}\dot{\SP}_{\;
  i}^{a}=\psi_{\;\left|i\right.}^{\left|a\right.},\] i.e. the time derivative of the 
displacement one--forms can now be represented by spatial derivatives of a potential
$\psi$ in the Newtonian approximation. The invariants of Eq. (\ref{eq:inInv})
similarly become
\begin{equation}
  \initial{\inI}:=\inI(\psi_{\;|j}^{|i}),\
  \initial{\inII}:=\inII(\psi_{\;|j}^{|i}),\
  \initial{\inIII}:=\inIII(\psi_{\;|j}^{|i}).
\label{eq:def-invariants-displacement}
\end{equation}
With this spatial geometrical limit we also send the general curved
space, on which we defined our average, to Euclidean space. Note that the existence of 
a vector field $f^a$ implies that the counterindex $a$ becomes a coordinate index of (now existing) global coordinates
in an embedding space. The Riemannian volume average automatically
corresponds to the Euclidean volume average over a flat domain
$\initial{\cD}$. (Note that sending the speed of light to infinity is only needed in order to change the
spacetime metric signature; since the backreaction model is spatial, and since we have eliminated all terms
that would contain the speed of light (the curvature), we do not have to send $c$ to infinity.)

Taking the spatial geometrical limit is thus sufficient to reduce the result, Eq.~(\ref{resultQ2}), 
to the Newtonian expression:
\begin{equation}
^{{\rm NZA}}\CQ_{\cD}=\frac{\dot{\xi}^{2}\;\left(\Upsilon_{1}+\xi\Upsilon_{2}+\xi^{2}\Upsilon_{3}\right)}{\left(1+\xi\laverage{\initial{\inI}}+\xi^{2}\laverage{\initial{\inII}}+\xi^{3}\laverage{\initial{\inIII}}\right)^{2}}\ ,\nonumber
\label{eq:Q-full-zel}
\end{equation}
with:
\begin{align}
  \begin{cases}
  & \Upsilon_{1}:=2\laverage{\initial{\inII}}-\frac{2}{3}\laverage{\initial{\inI}}^{2}\;\;,\qquad\qquad\\
  & \Upsilon_{2}:=6\laverage{\initial{\inIII}}-\frac{2}{3}\laverage{\initial{\inI}}\laverage{\initial{\inII}}\;\;,\;\;\\
  & \Upsilon_{3}:=2\laverage{\initial{\inI}}\laverage{\initial{\inIII}}-\frac{2}{3}\laverage{\initial{\inII}}^{2}\;\;.
\end{cases}
\end{align}

\subsection{Matching curvature expressions}
\label{sect:curv}

Another consistency check considers the scalar curvature in the RZA. There
are in principle three ways to calculate the intrinsic curvature that all agree for an exact solution.
The first way is to use the RZA metric and to calculate the curvature geometrically. The other two ways will be studied below; both relate the scalar curvature to extrinsic curvature invariants (note that these are relevant for this work, since we study kinematical backreaction being related to extrinsic curvature):
\begin{itemize}
\item [{\it a})] using a combination of the Hamilton constraint and
  the Raychaudhuri equation,
\item [{\it b})] using the integrability condition.
\end{itemize}
Thus the scalar curvature is a good tool to test the RZA: we expect that to the first
order the two results obtained by \textit{a}) and \textit{b}) are
the same. The expression for the way \textit{a}) has already been obtained as
Eq.~(\ref{eq:RZA-deviation-term}).
For the way \textit{b}) we express ${\cal W}_{\cD}$ through
Eq.~(\ref{eq:integrability-integral-GR}) as: 
\begin{equation} 
{\cal W}_{\cD}=\frac{1}{a_{\cD}^{2}}\left(c_{\initial{\cD}}-6k_{\initial{\cD}}-2\int_{{\initial
        t}}^{t}Q_{\cD}\frac{\rmd}{\rmd t'}a_{\cD}^{2}(t')\rmd t'\right)-Q_{\cD}\ .
\label{eq:WofQ}
\end{equation}
Using (\ref{eq:evolution-q}), (\ref{eq:backreaction-RZA-1}), the Friedmann equations for the
background scale factor, and 
\begin{equation} ^{\rm RZA}a_{\cD}=a(t)\aaverage{\frak J}^{1/3}\;,
\label{eq:a_D-vs-a}
\end{equation}
one can show after a long but not too technical calculation the following:
\begin{eqnarray}
  ^{{\rm RZA}}{\cal W}_{\cD}&=&-{\displaystyle \left\{ \frac{\aaverage{\ddot{{\frak J}}}}{\aaverage{{\frak J}}}+3\left(\frac{\ddot{\xi}}{\dot{\xi}}+4\frac{\dot{a}}{a}\right)\frac{\aaverage{\dot{{\frak J}}}}{\aaverage{{\frak J}}}\right\} }\nonumber \\
  +\frac{2}{a^{2}\aaverage{{\frak J}}^{2/3}}&\cdot&\intop_{\initial t}^{t}\left[\frac{8\pi G\rho_{\initial H}}{a \aaverage{{\frak J}}^{1/3}}q\left(t'\right)\frac{\rmd}{\rmd t'}\left(\frac{\aaverage{\dot{{\frak J}}}}{\dot{\xi}}\right)\right]\rmd t'\nonumber \\
  & & +6\left(\frac{k}{a^{2}}-\frac{k_{\initial{\cD}}}{a_{\cD}^{2}}\right)\;.
\label{eq:curv-normal-int-cond}
\end{eqnarray}
Using this expression for the averaged curvature assures the integrability condition (\ref{eq:integrability-GR}) to hold for a given 
kinematical backreaction model. In other words, this expression respects the conservation law for the combined action of kinematical
backreaction and averaged curvature (third equation of Eq.~(\ref{eq:effectivefriedmann}); note that the averaged rest mass is individually conserved). 
 
If we now compare Eq.~(\ref{eq:RZA-deviation-term}) and Eq.~(\ref{eq:curv-normal-int-cond}),
we find that the RZA is consistent if and only if the remaining integral
vanishes. After a change of integration variables this integral has
the form:
\begin{eqnarray}
  & {\displaystyle \int_{\initial{\xi}}^{\xi_{f}}{\displaystyle \frac{8\pi G\rho_{\initial H}}{a\left(\xi\right)\aaverage{{\frak J}}^{1/3}}\left(\xi+{\displaystyle \frac{q(\initial t)}{\dot{q}(\initial t)}}\right){\displaystyle \frac{\rmd^{2}}{\rmd\xi^{2}}{\displaystyle \aaverage{{\frak J}}}\rmd\xi}}=0\;.}%
\label{eq:RZA-consistency}
\end{eqnarray}
As it is zero for all values of $\xi_{f}$, already the integrand
has to vanish. Its first three factors are nonzero which means 
\begin{equation}
  \frac{\rmd^{2}}{\rmd\xi^{2}}{\displaystyle \aaverage{{\frak J}}}=0\;,\ \ \forall\xi\;.
\label{eq:conscondint}
\end{equation}
Using the definition of $\aaverage{{\frak J}}$ this implies that the
RZA approximation is consistent iff:
\begin{align}
  \begin{cases}
    a_\cD^3 = 1+\xi(t)\caverage{\initial{\inI}}\\
   \caverage{\initial{\inII}}=0=\caverage{\initial{\inIII}}\
    ,\end{cases}
\end{align}
which encodes the fact that it is strictly
speaking only a first--order scheme. Its success in the Newtonian case,
however, motivates its use in the relativistic case as well, despite
this result which otherwise stated reads: 
\begin{equation}
  {\displaystyle \int_{\initial{\xi}}^{\xi_{f}}{\displaystyle
      \frac{\xi+{\displaystyle \frac{q(\initial t)}{\dot{q}(\initial
            t)}}}{a\left(\xi\right)}{\displaystyle
        \frac{\caverage{\initial{\inII}}+3\xi\caverage{\initial{\inIII}}}{\caverage{{\frak J}}^{1/3}}\rmd\xi}}}=\mathcal{O}\left(\caverage{\initial{\inII}}\right)\;.
\end{equation}
This shows that the deviation from Eq.~(\ref{eq:RZA-deviation-term})
contains only terms of second and higher order in the initial conditions.

An equivalent way to check the consistency is to insert $^{{\rm RZA}}\CQ_{\cD}$ (Eq.~(\ref{resultQ2})), $^{{\rm RZA}}{\cal
  W}_{\cD}$ (Eq.~(\ref{resultW})) and $^{\rm RZA}a_{\cD}$(Eq.~(\ref{eq:a_D-vs-a})) directly into the integrability
condition Eq.~(\ref{eq:integrability-GR}). The result may be written
as \[
-\frac{6H_{0}\Omega_{m}\left(1+H_{0}\xi\right)}{a^{3}\left(t\right)\aaverage{{\frak J}}}\dot{\xi}\frac{\rmd^{2}}{\rmd\xi^{2}}\aaverage{{\frak J}}=0\;.\]
This can be interpreted as the amount by which the {}"closure'' of the
integrability condition of $^{{\rm RZA}}\CQ_{\cD}$ and $^{{\rm RZA}}{\cal W}_{\cD}$ fails.
As the prefactors are nonvanishing, we recover the condition of
Eq.~(\ref{eq:conscondint}).

\subsection{Self--consistency test in terms of the scale factor}

We have already seen above that the RZA delivers
only consistent expressions, if it is employed strictly as a first--order scheme. Therefore, an extrapolation as suggested in \cite{rza1},
bearing a number of advantages as discussed there,
is only an approximation for the terms at second and higher order.
It is therefore not surprising that, in addition to the different curvature expressions, we
also have two concurring definitions for the volume scale factor.  One
is the kinematical volume scale factor defined as the cubic root of the local volume deformation in the RZA,
Eq.~(\ref{eq:RZAJac}), $^{{\rm KIN}}a_{\cD}=\left(^{{\rm
      RZA}}{\rm J}\right)^{1/3}$.  The other one is $^{{\rm RZA}}a_{\cD}$,
calculated from the RZA of the backreaction term and using this latter as a source of the general equations governing the
average evolution, i.e. the
prescription of Subsect.~\ref{sect:scalefactor}.  Just like the
curvature expressions, they do not coincide for the RZA. For a
full n$^{\rm th}$--order calculation they are the same up to the given order. For the curvature we will see in Subsect.~\ref{sect:LTB} that
the expression derived from the backreaction term is more powerful and seems
more appropriate to capture the nonlinear evolution. With this insight we are going to use $^{{\rm RZA}}a_{\cD}$ in the concrete calculations presented further below.  
Fortunately, it will turn out that 
the possible error induced by instead calculating the kinematical volume scale factor is numerically rather
small: evaluating the quantity
\begin{equation}
  \epsilon_{a}=\frac{|^{{\rm RZA}}a_{\cD} \;-\; ^{{\rm KIN}}a_{\cD }|}{ ^{{\rm RZA}}a_{\cD}}\;,
\end{equation}
we found that
$\epsilon_{a}$ stayed well below 0.1 at all times, as long as
$^{{\rm RZA}}a_{\cD}$ did not approach zero and the domain's effective radius was larger than $16$ Mpc.

\section{Exact Subcases}
\label{sect:exact}

We are now going to study subclasses of exact solutions that are
contained in the above approximation. We will learn that (i) a subclass of the averaged plane--symmetric solutions,
(ii) a subclass of the averaged LTB solutions, as well as (iii) a subclass of the averaged Szekeres solutions are found by
suitably restricting the initial data. This fact demonstrates that the
investigated approximation interpolates between two kinematically
orthogonal exact solutions that are, however, unlike in the corresponding Newtonian model, 
subjected to constraints. This latter fact suggests that one may be able to 
construct more general backreaction models that include these cases in full generality.
However, we do not expect this to be possible for first--order deformations.

\subsection{The `Relativistic Zel'dovich Approximation' and plane
  collapse models}
\label{sect:plane-symmetry}

In Newtonian cosmology the `Zel'dovich approximation' is an {\em
  exact} three--di\-men\-sio\-nal solution to the Newtonian dynamics
of self--gra\-vi\-ta\-ting dust--matter for initial conditions with
$\inII\left(\theta_{ij}\right)=0=\inIII\left(\theta_{ij}\right)$ at
each trajectory \cite{buchert:class}. This {}"locally one--dimensional''
class of motions contains as a subcase the globally plane--symmetric
solution (see also \cite{buchertgoetz,barrowgoetz,silbergleit}). In the relativistic case one may
study the corresponding model using the plane--symmetric
ansatz for the line element,
\begin{equation}
  \rmd s^{2}=-\rmd t^{2}+a\left(t\right)^{2}\left(\rmd x^{2}+\rmd y^{2}+\left(1+P\left(z,t\right)\right)^{2}\rmd z^{2}\right)\;,%
\label{eq:planeSymMetric}
\end{equation}
which also has vanishing higher invariants of the peculiar expansion
tensor:
$\inII\left(\theta_{ij}\right)=0=\inIII\left(\theta_{ij}\right)$.  The
first invariant is nontrivial and reads
\begin{equation}
  \inI\left(\theta_{\;\;
      j}^{^{i}}\right)=\frac{\dot{P}\left(z,t\right)}{1+P\left(z,t\right)}\;.
\label{eq:planeInv}
\end{equation}
The equation determining the time evolution of $P\left(z,t\right)$ was
in the Newtonian case simply
\begin{equation}
  \dot{\Theta}+\Theta_{l}^{k}\Theta_{k}^{l}=-4\pi
  G\varrho+\Lambda\;,
\label{eq:RayNewton}
\end{equation}
which gave 
\begin{equation}
  \ddot{P}\left(z,t\right)+2\frac{\dot{a}}{a}\dot{P}\left(z,t\right)=4\pi
  G\varrho_{H}\left(P\left(z,t\right)-P_{0}\left(z\right)\right)\;.
\end{equation}
Hence, the Newtonian plane collapse had two solutions, e.g. for an EdS background,
\begin{equation}
  P\left(z,t\right)=P_{0}\left(z\right)+aC_{1}\left(z\right)+\frac{C_{2}\left(z\right)}{a^{3/2}}\;,
\end{equation}
a growing and a decaying one. In the relativistic case, however, there
are more constraints. In the Lagrange--Einstein--System of \cite{rza1},
also a link to the scalar curvature comes in: 
\begin{equation}
  \dot{\Theta}_{j}^{i}+\Theta\Theta_{j}^{i}=\left(4\pi
    G\varrho+\Lambda\right)\delta_{j}^{i}-\CR_{j}^{i}\;;
\label{eq:EinstThreeRicci}
\end{equation}
\begin{equation}
  \Theta^{2}-\Theta_{l}^{k}\Theta_{k}^{l}=16\pi G\varrho+2\Lambda-\CR\;.
\label{eq:HamConsRel}
\end{equation}
These two equations combined also give Eq.~(\ref{eq:RayNewton})
for the relativistic case. Additionally, however, they have to be
satisfied individually. As the plane--symmetric metric ansatz Eq.~(\ref{eq:planeSymMetric})
implies that $\CR_{\;j}^{i}=0$, the relativistic solution space is
not the same as the Newtonian one. For the Hamilton constraint Eq.~(\ref{eq:HamConsRel})
we find
\begin{equation}
  \frac{\dot{a}}{a}\dot{P}\left(z,t\right)=-4\pi G\varrho_{H}\left(P\left(z,t\right)-P_{0}\left(z\right)\right),\\
\end{equation}
which is now, for $\CR=0$, only a differential equation of first order
with the solution
\begin{equation}
  P\left(z,t\right)=P_{0}\left(z\right)+\frac{C\left(z\right)}{a^{3/2}}\;.
\label{eq:solPrel}
\end{equation}
This $P\left(z,t\right)$ also satisfies Eq.~(\ref{eq:EinstThreeRicci}),
but one of the solutions of the Newtonian case has disappeared.

From (\ref{eq:solPrel}) it follows that $P\left(z,t\right)=P_{0}\left(z\right)+q(t)C(z)$, with $q$ being a solution of (\ref{eq:evolution-q}). 
Inserting this result into (\ref{eq:planeInv}), and using
Eq.~(\ref{eq:backreaction-pelicular_tensor}), we find for the
backreaction term 
\begin{equation}
\CQ_{\cD}^{\;{\rm
      plane}}=-\frac{2}{3}\average{\inI\left(\theta_{\;\;
        j}^{^{i}}\right)}^{2}=-\frac{2}{3}\left(\frac{\dot{\xi}\caverage{\initial{\inI}}}{1+\xi\caverage{\initial{\inI}}}\right)^{2},
\end{equation}
where we recover $\xi\left(t\right)=\left(q\left(t\right)-q\left(t_0\right)\right)/\dot{q}\left(t_0\right)$, and where $\initial{\inI}=\inI(\theta_{\;\;
      j}^{^{i}}) (t_{i})$. Extracting the coframe from the metric (\ref{eq:planeSymMetric}) 
we find also that $\initial{\inI}=\inI\left(\dot{\mathscr P}_{\
    i}^{a}\left(z,t_{i}\right)\right)$, where $\dot{\mathscr P}_{\ i}^{a}\left(z,t_{i}\right)$ 
is the RZA perturbation. This shows that the plane--symmetric 
metric is, as in the Newtonian case, a particular exact solution that
is contained in the solutions of the RZA.  Note again, however, that
this solution in the RZA as well as for the plane--symmetric
metric does not have the growing mode that was present in the
Newtonian solution. This is due to the vanishing scalar curvature for cylindrical symmetry and
the relation to the Hamilton constraint that did not exist in the
Newtonian case. (Note that the integrability condition (\ref{eq:integrability-GR}) is satisfied by this solution.) 
This is another interesting example of a case in which
a class of Newtonian solutions may not automatically provide 
a solution of general relativity.

For negative $\caverage{\initial{\inI}}$, corresponding to overdense
regions (Eq.~(\ref{eq:Delta-I-1})), $\CQ_{\cD}^{\;{\rm plane}}$
is diverging at some time when $1+\xi(t)\caverage{\initial{\inI}}$
approaches zero, even though the solution is decaying. Our special
initial conditions imply a one--dimensional symmetry of
inhomogeneities on a three--dimensional background (cylindrical symmetry), 
and the diverging $\CQ_{\cD}^{\;{\rm plane}}$ is
supposed to mimic the highly anisotropic pancake collapse in the
three--dimensional situation.

The plane--symmetric case has also been considered in \cite{matarrese:plane}, but in the framework of a post--Newtonian approximation.
The evolution model discussed here does not need to invoke a post--Newtonian 
approximation, but this can be useful for a precise calculation of the initial invariants.

\subsection{The `Relativistic Zel'dovich Approximation' and spherical collapse models}
\label{sect:spherical-symmetry}

\subsubsection{The LTB solutions}
\label{sect:LTB}

We now investigate the Lema\^{i}tre--Tolman--Bondi (henceforth LTB)
solutions \cite{lemaitre-tolman-bondi}. The LTB model, written below in the time--synchronous
metric form,  is the general inhomogeneous spherically symmetric solution for irrotational dust. 
The spherical solution can be seen as
a superposition of infinitesimally thin homogeneous shells governed
by their own dynamics. In such a domain one can show (see
\cite{Enqvist} for a demonstration but with different notation, and the review \cite{sussman:review} for a comprehensive study of backreaction in the LTB solutions) that
the line element has the form: 
\begin{equation}
  \rmd s^{2}=-\rmd t^{2}+\frac{R'^{2}(t,r)}{1+2E(r)}\rmd r^{2}+R^{2}(t,r)\rmd \Omega^{2},
\end{equation}
$E$ being a free function of $r$ satisfying $E(r)>-1/2$; the prime
denotes partial differentiation with respect to $r$.\\
In this metric, the scalar parts of Einstein's field equations read:
\begin{equation}
  4\pi\rho(t,r)=\frac{M'(r)}{R'(t,r)R^{2}(t,r)},
\label{eq:rhoLTB}
\end{equation}
and 
\begin{equation}
  \frac{1}{2}\dot{R}^{2}(t,r)-\frac{GM(r)}{R(t,r)}=E(r),
\label{eq:ELTB}
\end{equation}
$M$ being another free function of $r$; the dot denotes partial time derivative.
Using the relation between the coefficients of the expansion tensor and the metric tensor,
\begin{equation}
  \Theta_{\; j}^{i}:=\frac{1}{2}g^{ik}\dot{g}_{kj},
\end{equation}
the averaged scalar invariants of the expansion tensor can be calculated:
\begin{eqnarray}
  \ltbaverage{\inI(\Theta_{\; j}^{i})} & = & \frac{4\pi}{V_{\rm LTB}}\int_{0}^{r_{\cD}}\frac{\partial_{r}\left(\dot{R}R^{2}\right)}{\sqrt{1+2E}}\rmd r\;;
\label{eq:invILTB}\\
  \ltbaverage{\inII(\Theta_{\; j}^{i})} & = & \frac{4\pi}{V_{\rm LTB}}\int_{0}^{r_{\cD}}\frac{\partial_{r}\left(\dot{R}^{2}R\right)}{\sqrt{1+2E}}\rmd r\;; \\
  \ltbaverage{\inIII(\Theta_{\; j}^{i})} & = & \frac{4\pi}{3V_{\rm LTB}}\int_{0}^{r_{\cD}}\frac{\partial_{r}\left(\dot{R}^{3}\right)}{\sqrt{1+2E}}\rmd r\;,
\label{eq:invIIILTB}
\end{eqnarray}
where the Riemannian volume of an LTB--domain is given by
\begin{equation}
  V_{\rm LTB}=\frac{4\pi}{3}\int_{0}^{r_{\cD}}\frac{\partial_{r}\left(R^{3}\right)}{\sqrt{1+2E}}\rmd r\;.
\end{equation}
The deviation from constant--curvature can also be averaged on a LTB--domain: 
\begin{equation}
  {\cal W}_{\rm LTB}=\ltbaverage{\CR} -\frac{6 k_{\initial\cD}V_{\initial{\rm LTB}}^{2/3}}{V_{\rm LTB}^{2/3}}\;,
  \label{eq:curvature-spherical}
  \end{equation} 
 with
 \begin{equation}
  \ltbaverage{\CR} = -\frac{16\pi}{V_{\rm LTB}}\int_{0}^{r_{\cD}}\frac{\partial_{r}\left(ER\right)}{\sqrt{1+2E}}\rmd r\ .
\end{equation}
There are two cases for which we find relations between the invariants
without having to solve the system (\ref{eq:rhoLTB}) and (\ref{eq:ELTB})
explicitly: the first one is a separable solution $R\left(t,r\right)$ of the
form\[R\left(t,r\right)=f\left(t\right)\cdot g\left(r\right)\;.\]
The second one is the case of an LTB domain where $E(r)=E$.
The restriction $E=const.$ corresponds to self--similar LTB solutions if we require at the same time that the function $M(r) \propto r$ \cite{sussman}.

In both cases, one can show for
$R\left(t,0\right)=0$: 
\begin{eqnarray}
  & \ltbaverage{\inII(\Theta_{\; j}^{i})}={\displaystyle \frac{1}{3}\ltbaverage{\inI(\Theta_{\; j}^{i})}^{2}\;,}\nonumber \\
  & \ltbaverage{\inIII(\Theta_{\; j}^{i})}={\displaystyle
    \frac{1}{27}\ltbaverage{\inI(\Theta_{\;
        j}^{i})}^{3}\;.}
\label{eq:spherical-I-III}
\end{eqnarray}
Combining these terms in the backreaction $\CQ_{LTB}$ given by
Eq.~(\ref{eq:Q-GR}), we get for a spherically symmetric $E =const.$--domain or
a separable $R\left(t,r\right)$: 
\begin{equation}
\CQ_{\rm LTB}=0\ \;,\
  \ {\cal W}_{\rm LTB}=0\;,
\end{equation}
where the result ${\cal W}_{\rm LTB}=0$ follows from $\CQ_{\rm LTB}=0$ by the integrability
condition (\ref{eq:integrability-integral-GR}) and its definition
(\ref{eq:avpeccurv}). We here generalize a result obtained in
\cite{curvatureLTB} to a nonflat domain for a special case. 

Note that, using the expression for the curvature of Eq.~(\ref{resultW}), we find that the result is not zero for
the invariants Eq.~(\ref{eq:spherical-I-III}).  This means that this
expression is not yet containing the correct second--order
contributions. This is one important reason for why we use in the evaluation of the importance of
$\CW_{\cD}$ in Section~\ref{sect:evolution-density-par} the numerically
integrated expression, starting from the $\CQ_{\cD}$ of Eq.~(\ref{resultQ2}).

In the case $E= const.$, Eq.~(\ref{eq:curvature-spherical}) and
(\ref{eq:avpeccurv}) give for the averaged scalar curvature
\begin{equation} 
\ltbaverage\CR=-
    \frac{12E}{R^{2}(r_{\cD})}\;.
\label{eq:const-curv-result}
\end{equation}
With
${\cal W}_{\rm LTB}=0$, one can express $k_{\initial{\cD}}$ as a function
of $E$:
\begin{equation}
  k_{\initial{\cD}}=-\frac{2E}{R^{2}(\initial t,r_{\cD})}<\frac{1}{R^{2}(\initial t,r_{\cD})}\;.
\label{eq:const-curv-k-vs-E}
\end{equation}
As $R$ is a growing function of $r_{\cD}$, $k_{\initial{\cD}}$
becomes smaller when we increase the averaging domain.

\subsubsection{The Szekeres solutions}
\label{sect:szekeres}

A possible generalization of the LTB models is the quasispherical
Szekeres model that has additional anisotropies which destroy the
spherical symmetry of the LTB models \cite{bolejko:szekeres}. The line element reads:
\begin{equation}
\rmd s^{2}=-\rmd t^{2}+X^{2}\rmd r^{2}+Y^{2}\left(\rmd x^{2}+ \rmd y^{2}\right)\;,
\end{equation}
and with $\zeta=x+\imath y$\;\;;\;\;$\bar{\zeta}=x-\imath y$, we have
$$
X=\frac{\CE\left(r,\zeta,\bar{\zeta}\right)Y^{\prime}\left(t,r,\zeta,\bar{\zeta}\right)}{\sqrt{\epsilon-k\left(r\right)}}\;;\;
Y=\frac{R\left(t,r\right)}{\CE\left(r,\zeta,\bar{\zeta}\right)}\;;
$$
$$
\CE\left(r,\zeta,\bar{\zeta}\right)=a\left(r\right)\zeta\bar{\zeta}+b\left(r\right)\zeta+c\left(r\right)\bar{\zeta}+d\left(r\right)\;;\;\epsilon=0,\pm1\, ,
$$
where a prime denotes $\partial_{r}$. We consider
only the quasispherical model $\epsilon=1$ in which case, for
$\CE^{\prime}=0$, it simply reproduces the LTB model. But even for the
general case $\CE^{\prime}\ne0$ it has been shown in
\cite{bolejko:szekeres} that the invariants take the form of
Eqs. (\ref{eq:invILTB}) to (\ref{eq:invIIILTB}) with
$k\left(r\right)=-2E\left(r\right)$. Therefore also the flat
quasispherical Szekeres model has no backreaction:
\begin{equation}
  \CQ_{\rm Sz}=0\ \;,\ \ {\cal W}_{\rm Sz}=0\;.
\end{equation}
We considered only this example for the flat case, but the other flat cases should be accordingly solvable. However, in the really interesting cases, i.e. for arbitrary curvature, it is highly difficult to find a general statement, as it is also the case for the LTB model. For further work on the Szekeres model see \cite{meures.bruni,krasinski:szekeres}, and in relation to observations \cite{szekeres:obs1,szekeres:obs2,szekeres:obs3}.

\subsubsection{GR theorems corresponding to Newton's `Iron Sphere Theorem'}

An interesting property arises, if we look at situations where the kinematical backreaction term vanishes. 
From the averaged equations it then follows that, e.g. a flat spherically symmetric but inhomogeneous domain, cut
out of a homogeneous FLRW model, has no influence on the kinematical expansion properties. Such a result is known as 
Newton's `Iron Sphere Theorem' in a Newtonian model. 
Let us now look at situations where 
the backreaction term of the RZA approximation vanishes. Using it in the form of Eq.~(\ref{resultQ2}), one can
show the following propositions:\\
\begin{propos1} 
\label{propos1} \[ ^{{\rm RZA}}{\cal
    Q_{D}}=0\Leftrightarrow\left\{ \begin{array}{ll}
      \caverage{\initial{\inII}}={\displaystyle \frac{1}{3}\caverage{\initial{\inI}}^{2}\;}\\
      \\\caverage{\initial{\inIII}}={\displaystyle
        \frac{1}{27}\caverage{\initial{\inI}}^{3}\;.}\end{array}\right.\]
\end{propos1}\textit{}\\
\textit{Proof.} see Appendix~\ref{sect:AppendixProp}.\\
\begin{propos2} 
\label{propos2} \[ ^{{\rm RZA}}{\cal
    Q_{D}}=0\Leftrightarrow\left\{ \begin{array}{ll}
      \average{^{\rm RZA}{\rm II}\left(\Theta_{\ j}^{i}\right)}={\displaystyle \frac{1}{3}\average{^{\rm RZA}{\rm I}\left(\Theta_{\ j}^{i}\right)}^{2}\;}\\
      \\\average{^{\rm RZA}{\rm III}\left(\Theta_{\
            j}^{i}\right)}={\displaystyle
        \frac{1}{27}\average{^{\rm RZA}{\rm I}\left(\Theta_{\
              j}^{i}\right)}^{3}\;.}\end{array}\right.\]
\end{propos2}\textit{}\\
\textit{Proof.} see Appendix~\ref{sect:AppendixProp}.

This means that, for a flat background, the RZA correctly reproduces
the average evolution of the LTB matter shells, without knowing the
specific distribution of matter $M\left(r\right)$. This corresponds to the 
`Iron Sphere Theorem' of Newton in the relativistic case (compare also \cite{curvatureLTB}), 
and it is in the spirit of Birkhoff's and almost--Birkhoff theorems, but for nonvacuum spacetimes 
(see the recent papers \cite{ellis:birkhoff1,ellis:birkhoff2} and references therein).

In \cite{bks} it was shown that, in the
Newtonian case, the averaged Zel'dovich approximation also describes the
behavior of a spherical matter distribution. Here we find the
analogous property for the flat LTB and flat Szekeres metrics.

To summarize one may say that the RZA approximation interpolates between the exact
GR solutions of a plane--symmetric metric, the flat LTB and the flat
Szekeres metrics, which adds reliability to the employed extrapolation of a strictly first--order scheme and its use
within the exact averaging framework.

\section{The evolution of cosmological parameters}
\label{sect:evolution-cosmo}

The backreaction term itself decays in the averaged "Relativistic Zel'dovich
Approximation''.  However, to quantify the deviations from the
behavior of the scale factor and especially its time derivatives in the standard model, the different
strengths of the sources in the generalized Friedmann equation have to
be compared. As the matter source term decays faster than the
backreaction term, the influence of this latter grows. To evaluate the
importance of this growth in the standard cosmological picture,
starting with a nearly homogeneous and isotropic initial state, we
only need to determine the magnitude of the three invariants of the
perturbation one--forms $\dot{\mathscr P}_{\ i}^{a}$.  The time evolution 
of $^{{\rm RZA}}\CQ_{\cD}$ is then determined by
Eq.~(\ref{resultQ2}).  It then also determines the evolution of all
other cosmic parameters via the averages of the Hamilton constraint,
Eq.~(\ref{eq:hamiltonconstraint}), and Raychaudhuri's equation,
Eq.~(\ref{eq:expansion-law-GR}). The context of the standard scenario
implies that the calculation of the initial values is performed in a
universe model that is close to spatially flat, looking back to a history that is 
identical to the standard model.  If we additionally neglect
tensor modes, this means that we are in the limit described in
Sect.~\ref{sect:newtonlimit}. We therefore assume that we can use the
values of $\laverage{\initial{\inI}}$, $\laverage{\initial{\inII}}$
and $\laverage{\initial{\inIII}}$ in \cite{bks} as our initial
conditions for the averaged invariants of $\dot{\mathscr P}_{\ i}^{a}$ to a very good approximation.  In
formal analogy of Eq.~(\ref{resultQ2}) to the corresponding expression of \cite{bks}, this
then implies that many results of \cite{bks} carry over to
the RZA context. There are, however, new phenomena that emerge
due to the fact that, unlike in the Newtonian ZA of \cite{bks}, geometry is a dynamical variable and
the RZA develops nonvanishing scalar curvature. In this section, we
will therefore comment on which results of \cite{bks} remain valid and
discuss where the GR description brings in new phenomena.
As we emphasized already in the introduction, an emerging curvature is key to the new interpretation of
the dark components in the standard model.

\subsection{Calculation of the scale factor $a_{\cD}$}
\label{sect:scalefactor}

We calculate the average scale factor $^{{\rm
    RZA}}a_{\cD}$ as in \cite{bks} directly by integrating the averaged
Raychaudhuri Equation Eq.~(\ref{eq:expansion-law-GR}). The input is
the average density
\begin{equation}
  \average{\varrho}=\frac{a^{3}}{a_{\cD}^{3}}\varrho_{H}\left(1+\laverage{\delta\left(\initial t\right)}\right)=\frac{a^{3}}{a_{\cD}^{3}}\varrho_{H}\left(1-\laverage{\initial{\inI}}\right),
\label{eq:Delta-I-1}
\end{equation}
and the RZA  backreaction model, Eq.~(\ref{resultQ2}). The initial
conditions for $^{{\rm RZA}}a_{\cD}$ are $a_{\initial{\cD}}=1$
and ${\dot{a}_{\cD}}\left(\initial t\right)={\dot{a}}\left(\initial t\right)\left(1+\frac{1}{3}\laverage{\initial{\inI}}\right)$.

\subsubsection{Statistics of initial conditions}

As discussed, we use for the numerical evaluation the
initial values given in \cite{bks}. As was shown there, the expected
values of the initial invariants are zero,
\begin{equation}
  \BE\left[\laverage{\initial{\inI}}\right]=\BE\left[\laverage{\initial{\inII}}\right]=\BE\left[\laverage{\initial{\inIII}}\right]=0.
\label{eq:I=00003D0=00003DII=00003D0=00003DIII}
\end{equation}
$\BE\left[\dots\right]$ denotes the ensemble expectation value over many realizations of universe models. 
However, for a specific domain, any of the volume--averaged invariants
may be positive or negative. These invariants fluctuate with the variance,
e.g., $\sigma_{\inI}^{2}(R)=\BE\left[\baverage{\initial{\inI}}^{2}\right]$.
In our calculation of the time evolution of $a_{\cD}(t)$, we consider
one--$\sigma$ fluctuations of the averaged invariants for spherical
domains of radius $R$, e.g.\ $\baverage{\initial{\inI}}=\pm\sigma_{\inI}\left(R\right)$.
\cite{bks} showed how these fluctuations are linked to the matter
power spectrum. As indicated, $\sigma_{\inI}(R)$ will explicitly
depend on the radius of the initial domain, but implicitly also on the
shape of the power spectrum. \cite{bks} used a standard
CDM power spectrum normalized to $\sigma_{8}=1$ and an $h$ of $h=0.5$.
In addition to the EdS background considered there, we will also present
some results for a standard $\Lambda$CDM background with $\Omega_{\Lambda}\approx0.73$,
$h=0.7$ and $\sigma_{8}=0.8$.

Although the possible initial conditions have a rich structure due to the fact that we can choose any combination of the signs of the three initial invariants,
we choose for the presentation equal signs for all the invariants, i.e. for expanding, underdense domains, $\sigma_{\inI} > 0$, all initial invariants are taken to be positive, and for collapsing, overdense domains, $\sigma_{\inI} < 0$, all initial invariants are taken to be negative. (A showcase of mixed signs has been analyzed in \cite{bks}.)

It is important to recognize that our choice of using one--$\sigma$
fluctuations in the invariants does not imply that the fluctuations of
other parameters constructed from them are then also one--$\sigma$
fluctuations of these parameters. For the initial value of $\CQ_{\cD}$
for example a one--$\sigma$ fluctuation of the invariants is related to a
one--$\sigma$ fluctuation of $\CQ_{\cD}$ by
\begin{equation}
  \sigma^{2}\left[^{{\rm RZA}}\CQ_{\initial{\cD}}\right]=4\sigma_{\inII}^{2}\left(R\right)-\frac{8}{3}\BE\left[\laverage{\initial{\inI}}^{2}\laverage{\initial{\inII}}\right]
+\frac{8}{9}\sigma_{\inI}^{4}\left(R\right).
\label{eq:VarQ-corr}
\end{equation}
On the other hand, calculating $^{{\rm RZA}}\CQ_{\initial{\cD}}$
with one--$\sigma$ fluctuations yields
\begin{equation} 
^{{\rm RZA}}\CQ_{\initial{\cD}}=2\sigma_{\inII}\left(R\right)-\frac{2}{3}\sigma_{\inI}^{2}\left(R\right).
\end{equation}
In an abuse of language we will nevertheless speak of one--$\sigma$
fluctuations in the following, but one should keep in mind that we
mean those of the initial invariants (except for Fig.~\ref{fig:VarQ}). 
For the values of $^{{\rm RZA}}\CQ_{\initial{\cD}}$, in the case of  large
values of $R$, the two prescriptions coincide as
$\sigma_{\inI}^{2}\left(R\right)$ drops off faster than
$\sigma_{\inII}\left(R\right)$.

\subsubsection{Scale--dependence of initial conditions}

From the very definition of the averaged parameters it seems natural
that all averaged quantities would be scale dependent and decay
with a growing domain $\cD$. That this is not necessarily the case shows
the example of the initial expansion-- and shear--fluctuations.  In
the RZA they are given by
\begin{equation}
  \average{\theta^{2}}-\average{\theta}^{2}=\average{\inI\left(\dot{\mathscr
        P}_{\ i}^{a}\right)^{2}}-\average{\inI\left(\dot{\mathscr
        P}_{\ i}^{a}\right)}^{2}\;;
\end{equation}
\begin{equation}
  \average{\sigma^{2}}=\frac{1}{3}\average{\inI\left(\dot{\mathscr P}_{\ i}^{a}\right)^{2}}-\average{\inII\left(\dot{\mathscr P}_{\ i}^{a}\right)}.
\end{equation}
Using the acceleration equation in terms of coframes of \cite{rza1},
one can show that for the RZA we have $\inI\left(\dot{\mathscr P}_{\ i}^{a}\right)=-\delta$,
where $\delta$ is the local density contrast. This means that also
in the RZA, as already in the Newtonian case, the expectation values
of expansion-- and shear--fluctuations are no longer scale dependent.
They are rather given by
\begin{equation}
  \BE\left[\average{\theta^{2}}-\average{\theta}^{2}\right]=\initial H^{2}\left(\int_{\mathbb{R}^{3}}\rmd^{3}k\initial P\left(k\right)-\sigma_{\inI}^{2}\left(R\right)\right);
\end{equation}
\begin{equation}
  \BE\left[\average{\sigma^{2}}\right]=\frac{1}{3}\initial H^{2}\int_{\mathbb{R}^{3}}\rmd^{3}k\initial P\left(k\right),
\end{equation}
where we again made use of the assumed approximate flatness of the initial universe model, necessary to employ
the Fourier transformation. Interestingly, only part of
the expected expansion fluctuations still contains the information
on the domain $\cD$. The other part and the shear are domain independent.
Calculating the value of this integral may be used to estimate the
importance of backreaction, if the shear fluctuations were negligible.
To this end we calculate 
\begin{equation}
  \BE\left[^{{\rm RZA}}\Omega_{\CQ,trunc}^{\cD_{0}}\right]=-\frac{1}{9H_{\cD_{0}}^{2}}\BE\left[\averageDN{\theta^{2}}-\averageDN{\theta}^{2}\right],
\end{equation}

\noindent
where $0$ stands for today and we evolved the model from the
initial time up to today with the leading $a^{-1}$ mode of Eq.~(\ref{resultQ2})
only. A quantitative estimate with an exponential IR cutoff at the
Hubble scale and UV cutoff at $1$kpc then yields $\BE\left[^{{\rm RZA}}\Omega_{\CQ,trunc}^{\cD_{0}}\right]\approx0.73$.
This illustrates the well--known fact that the expansion-- and shear--fluctuations
by themselves are important even in a perturbative framework. In the
backreaction term, however, they combine in a way that leaves only the
domain--dependent contribution in the second term of the expansion
fluctuations. In the Newtonian framework this is expected by the fact
that $\CQ_{\cD}$ can be written as a surface term. In GR it is not
a necessity, but in the linear RZA the cancellation is still effective.
For higher orders, however, this is no longer true and \cite{clarkson}
reported the survival of a domain--independent contribution to $\CQ_{\cD}$
at second order.

\begin{figure}
  \includegraphics[width=0.5\textwidth]{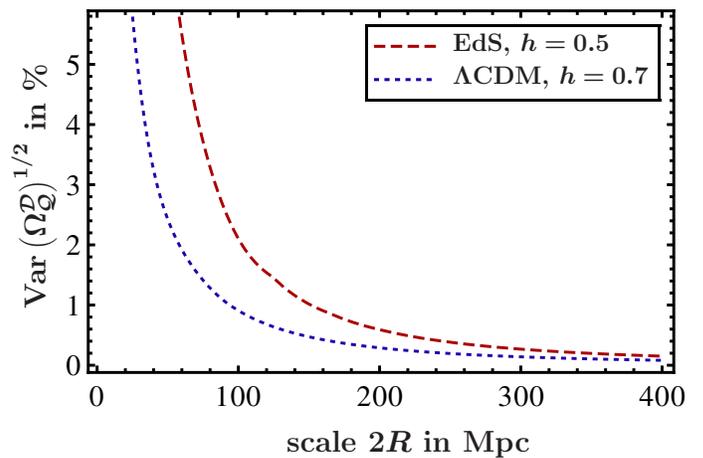}
  \caption{One--$\sigma$ fluctuations of $\Omega_{\CQ}^{\cD_{0}}$ as a function of scale for (i) an
    EdS background ($h=0.5$, $\sigma_{8}=1$) and (ii) for a $\Lambda$CDM background
    ($\Omega_{\Lambda}\approx0.73$, $h=0.7$,
    $\sigma_{8}=0.8$).
\label{fig:VarQ}}
\end{figure}

To close this section we calculate, as an illustration of the 
scale dependence of the parameters, the backreaction term
$\CQ_{\cD}$. It will imprint its scale dependence on the other
parameters and is therefore particularly interesting. To get a feeling
for the magnitude of the values, we evolve it again with the $a^{-1}$
mode until today and normalize it with $H_{\cD_{0}}^{2}$ to get
$\Omega_{\CQ}^{\cD_{0}}$.  The result is shown in
Fig.~\ref{fig:VarQ}. We plot the one--$\sigma$ fluctuation of
$\Omega_{\CQ}^{\cD_{0}}$ with the correct $\sigma-$interpretation of
Eq.~(\ref{resultQ2}). The result shows that only below the assumed
"almost homogeneity scale" of about $400$ Mpc the (with $a^{-1}$ upscaled) backreaction term 
begins to enter the range of a percent contribution. For larger scales it is clearly below this value, while for smaller scales
the nonlinear terms in the backreaction functional count in. This explains
that in the following all parameters converge to their background
values for large $\cD$. It is here where our model faces its strongest restriction, since we
neglect any interaction of structure with the assumed background model.

\subsection{Time evolution of cosmic parameters}

Having discussed in the previous section, how we fix our initial
conditions and determine the scale factor from them, we will now
present and discuss the results.

\subsubsection{Evolution of the scale factor $a_{\cD}$, the Hubble--
  and deceleration--parameters}
\label{sect:evolution-HD-qD}

\begin{figure*}
  \includegraphics[width=0.47\textwidth]{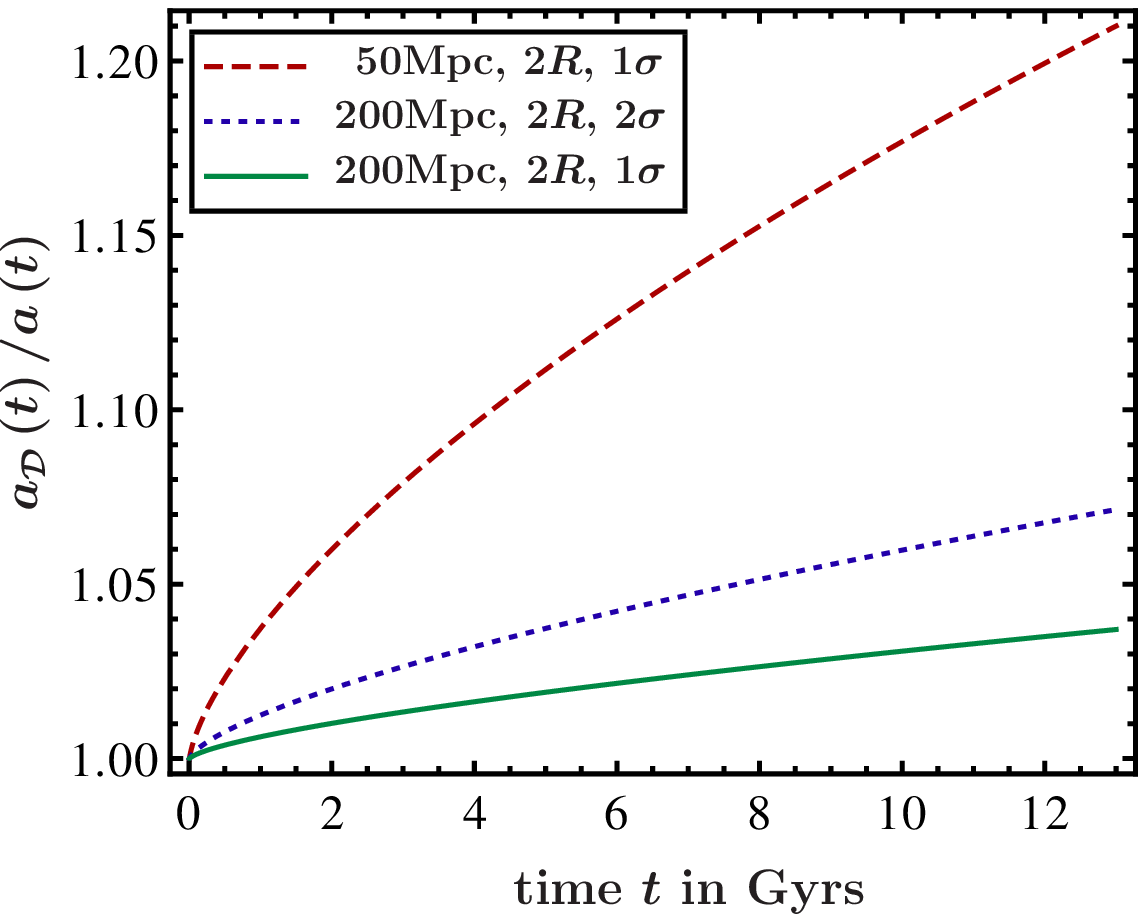}\hspace{20pt}
  \includegraphics[width=0.47\textwidth]{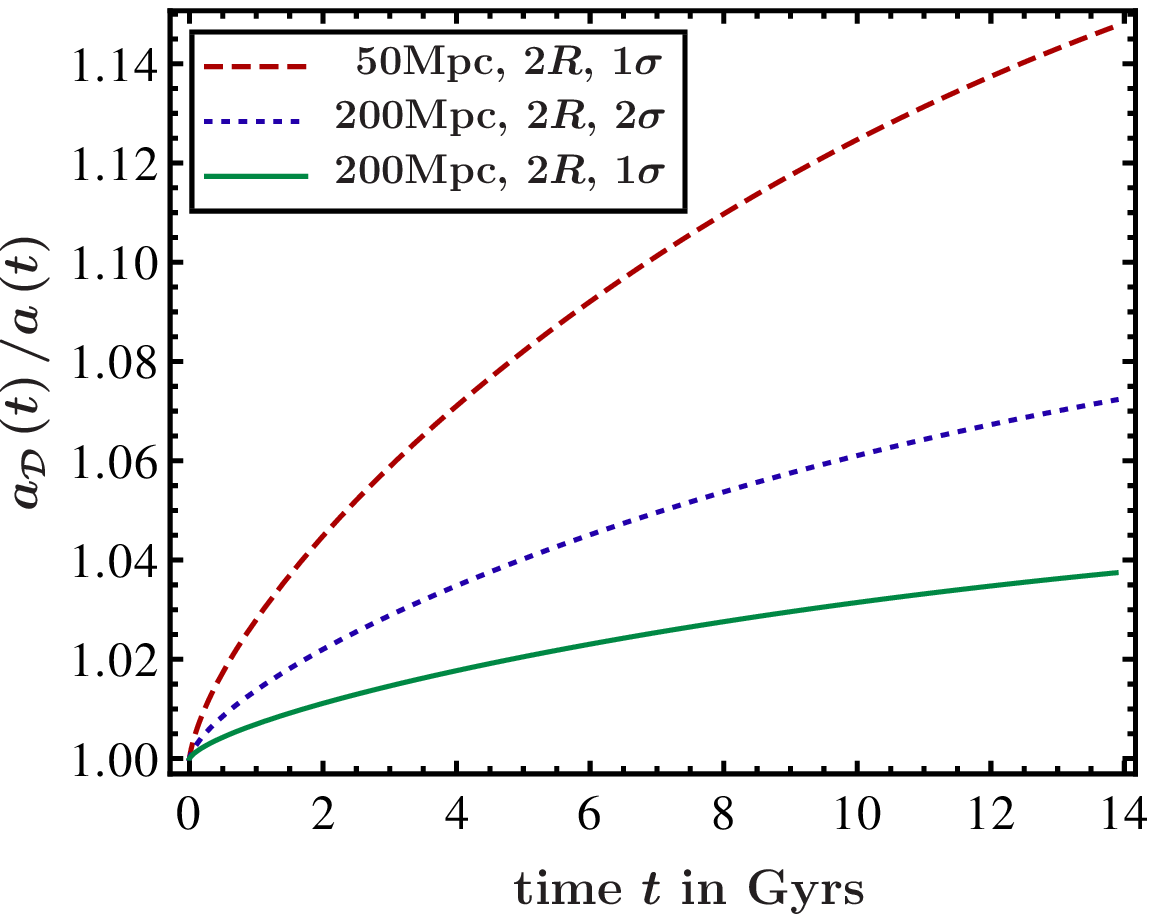}
    \caption{Evolution of the volume scale factor, normalized by the background scale factor, on typical domains of $50$ and $200$ Mpc,
    and on a domain that is a $2-$sigma fluctuation in the initial conditions on $200$ Mpc for comparison.  
    Left: EdS background with
    $\Omega_{m}=1$ ($h=0.5$, $\sigma_{8}=1$). Right: $\Lambda$CDM background with $\Omega_{m}=0.27$ ($h=0.7$,
    $\sigma_{8}=0.8$). 
    \label{fig:scalefactors}}
\end{figure*}

\begin{figure*}
  \includegraphics[width=0.45\textwidth]{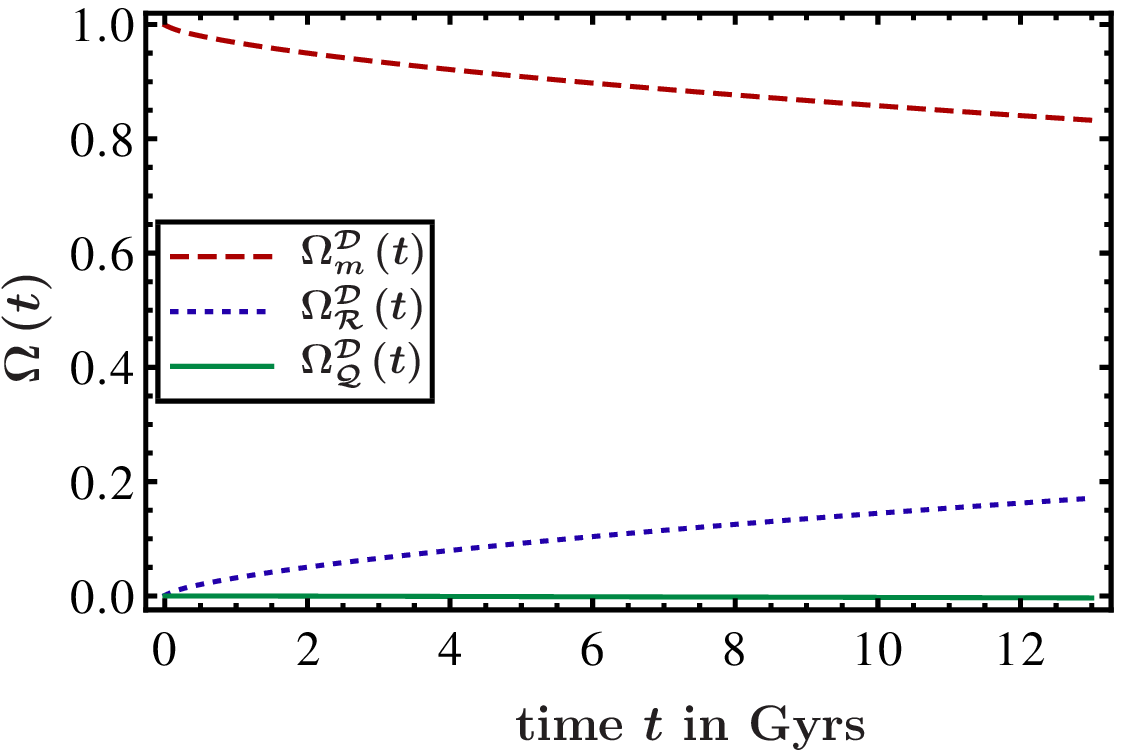}\hspace{20pt}
  \includegraphics[width=0.45\textwidth]{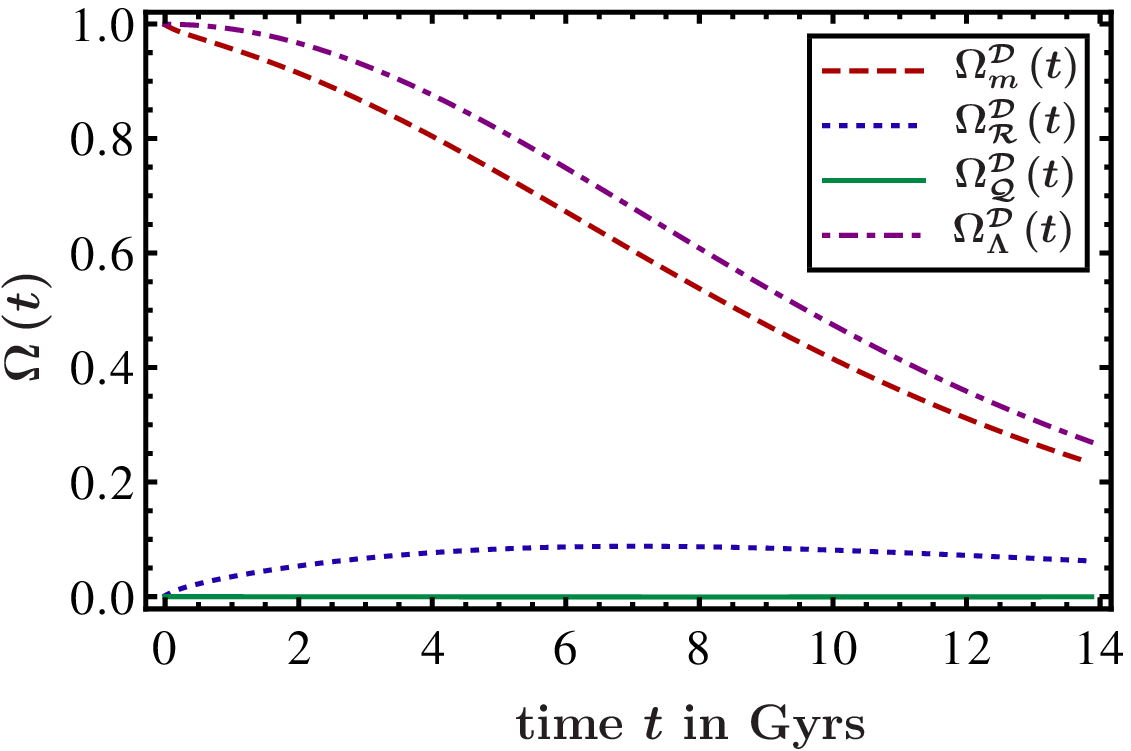}
    \caption{Evolution
    of the domain dependent cosmological parameters of Eq.~(\ref{eq:cospar})
    with cosmic time. One background is the EdS model with
    $\Omega_{m}=1$ ($h=0.5$, $\sigma_{8}=1$) (left), the other one the
    $\Lambda$CDM model with $\Omega_{m}=0.27$ ($h=0.7$,
    $\sigma_{8}=0.8$) (Right: the background density parameter is plotted here as the upper curve). 
    The figure shows values for an expanding underdense domain of
    $200$ Mpc effective diameter  with one--$\sigma$ fluctuations of the initial
    invariants of the perturbation one--form.
    This figure confirms the findings of \cite{bks} reporting substantial deviations from
    the background of e.g. the matter density parameter, while the quantitative importance of the backreaction parameter is 
    seemingly negligible. The new interpretation in the GR context is mirrored by the curvature density parameter that has to 
    compensate (in the case of an EdS background without a `Dark Energy' component) the large deviations in the matter density parameter.
    \label{fig:cospar100}}
\end{figure*}

As mentioned, the evolution of the dynamical quantities $a_{\cD}$,
$H_{\cD}={\dot{a}}_{\cD}/a_{\cD}$ and
$q_{\cD}=-(\ddot{a}_{\cD} / a_{\cD}) / H_{\cD}^{2}$ turns out to
coincide with those of \cite{bks} for our choice of initial
conditions. This means that the partially drastic deviation from the
background values for these quantities, that has been found in
\cite{bks}, also occurs in the RZA. In the case of the volume deceleration
parameter $q_{\cD}$ for one--$\sigma$ fluctuations this leads to
deviations of $30$\% on a scale of $200$ Mpc. 

The effects on $a_{\cD}$
and $H_{\cD}$ are smaller, but may also become important for special
regions, that are more than one--$\sigma$ away from the background.
In the relativistic framework the deviations of the volume scale factor $a_\cD$ from
the background scale factor $a(t)$ can also be interpreted as giving the strength of
metrical perturbations, since the calculation of the volume just involves the metric
determinant, and not higher derivatives of the metric. Taking the cube of this deviation 
gives us a typical strength for the volume fluctuation, e.g. for one--sigma fluctuations on 
the scale of $200$ Mpc we find a $12$ percent effect,  
see Fig.~\ref{fig:scalefactors} for the evolution of the scale factors. This also means
that the influence of metrical perturbations is not necessarily small, if their averaged effect is considered (see the related discussions in \cite{ellisFOCUS,rasanen:lightpropagation,buchertrasanen}). However, it is still
subdominant compared with the overall backreaction effect that depends on second derivatives of the metric.

In the next subsection we will discuss the results for the density parameters that are relevant for the interpretation of the energy budget of the
Universe and its relation to the `dark components' in the standard model. Recall that we quantify fluctuations on Lagrangian domains in this paper;
for a quantification of fluctuations of cosmic parameters on domains that correspond to actual observational geometries see \cite{variance}.

\subsubsection{Evolution of the density parameters}
\label{sect:evolution-density-par}

\begin{figure*}
  \includegraphics[width=0.47\textwidth]{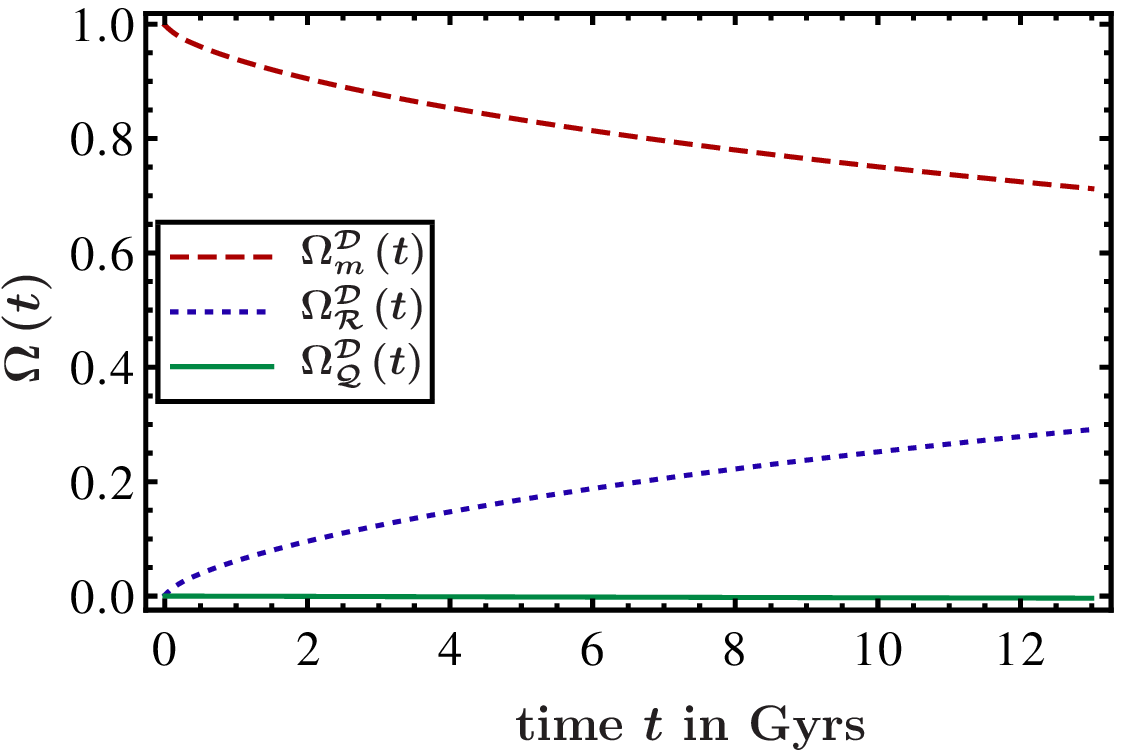}\hspace{20pt}
  \includegraphics[width=0.47\textwidth]{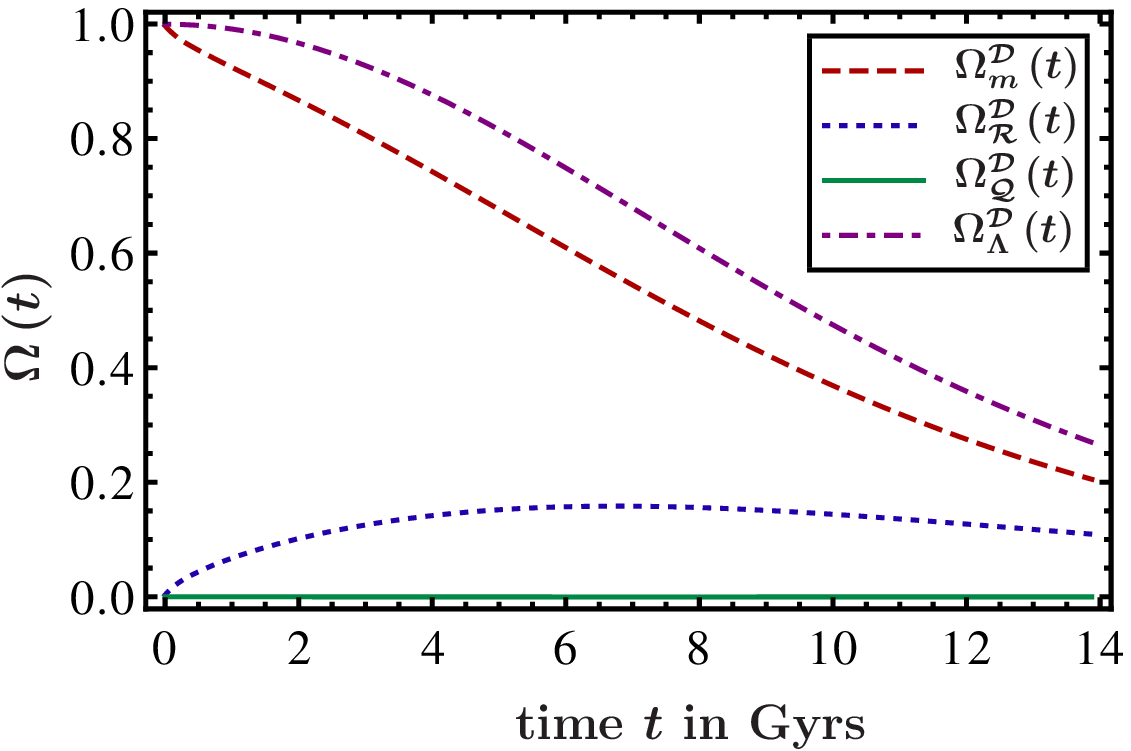}
    \caption{
    This figure corresponds to Fig.~\ref{fig:cospar100}, but shows the corresponding values for an expanding domain of
    two--$\sigma$ fluctuations of the initial
    invariants of the perturbation one--form.
\label{fig:cospar100sig2}}
\end{figure*}

\begin{figure*}
  \includegraphics[width=0.47\textwidth]{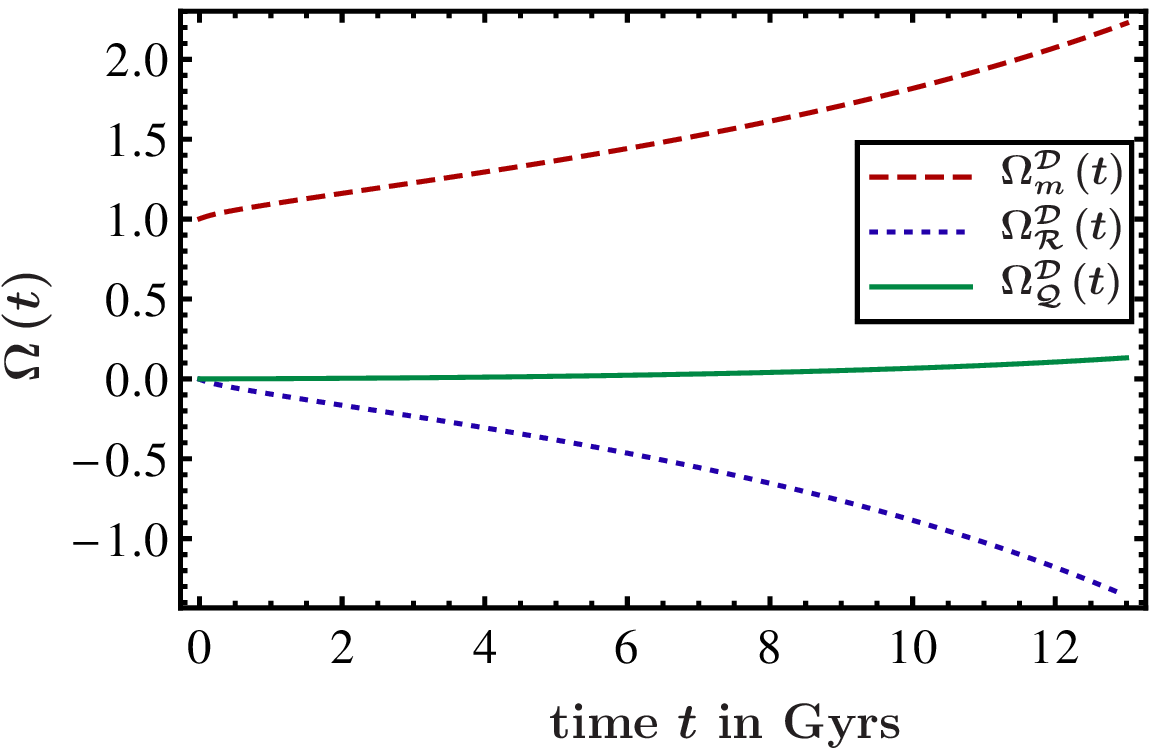}\hspace{20pt}
  \includegraphics[width=0.47\textwidth]{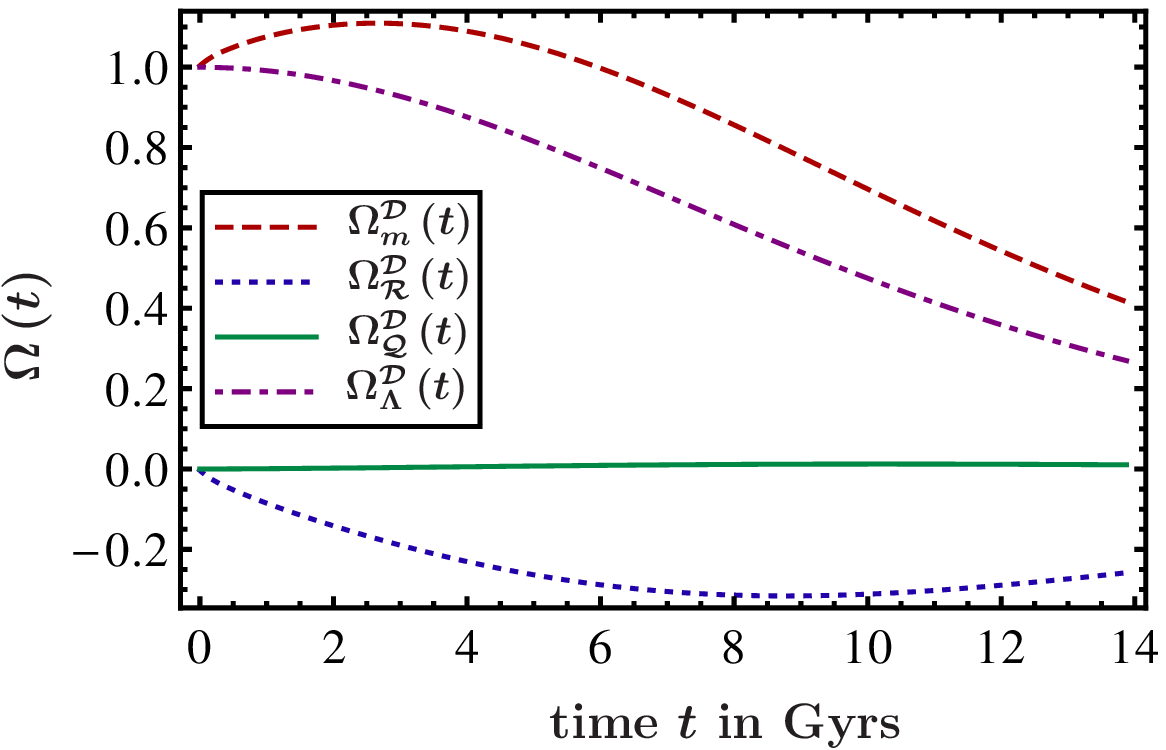}
    \caption{Evolution
    of the domain dependent cosmological parameters of Eq.~(\ref{eq:cospar})
    with cosmic time. One background is the EdS model with
    $\Omega_{m}=1$ ($h=0.5$, $\sigma_{8}=1$) (left), the other one the
    $\Lambda$CDM model with $\Omega_{m}=0.27$ ($h=0.7$,
    $\sigma_{8}=0.8$) (Right: the background density parameter is plotted here as the lower curve). 
    The figure shows values for a collapsing overdense domain of
    $100$ Mpc effective diameter with one--$\sigma$ fluctuations of the initial
    invariants of the perturbation one--form. 
\label{fig:cospar50}}
\end{figure*}

\begin{figure*}
  \includegraphics[width=0.45\textwidth]{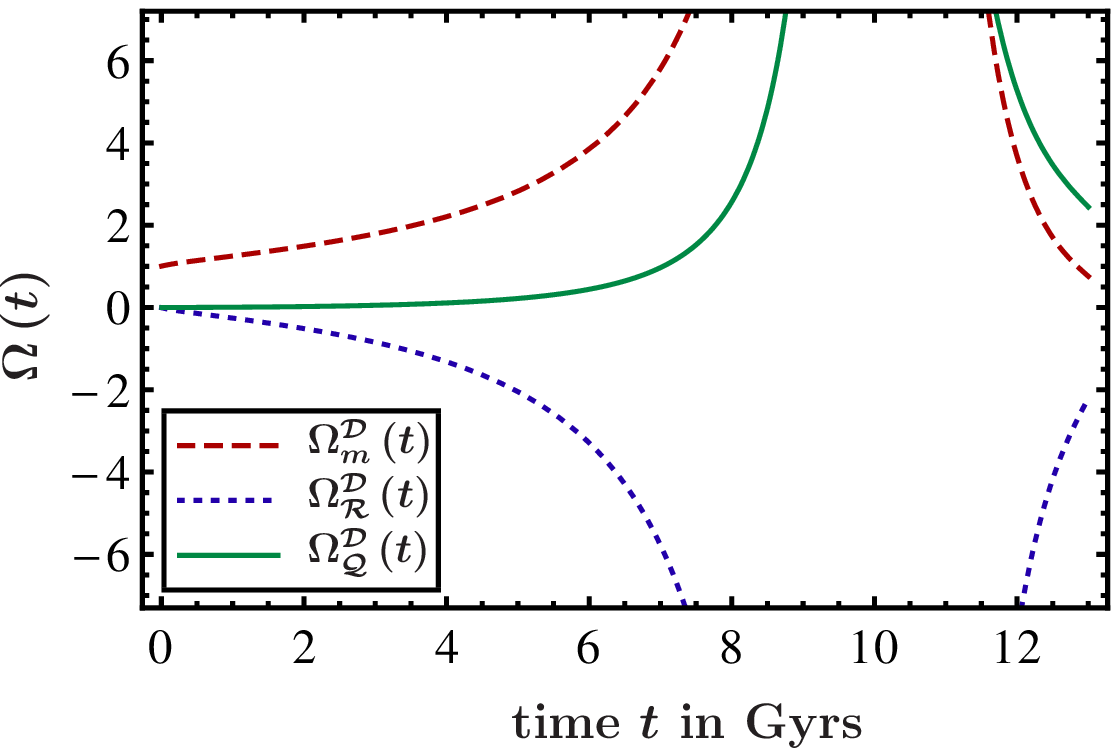}\hspace{20pt}
  \includegraphics[width=0.45\textwidth]{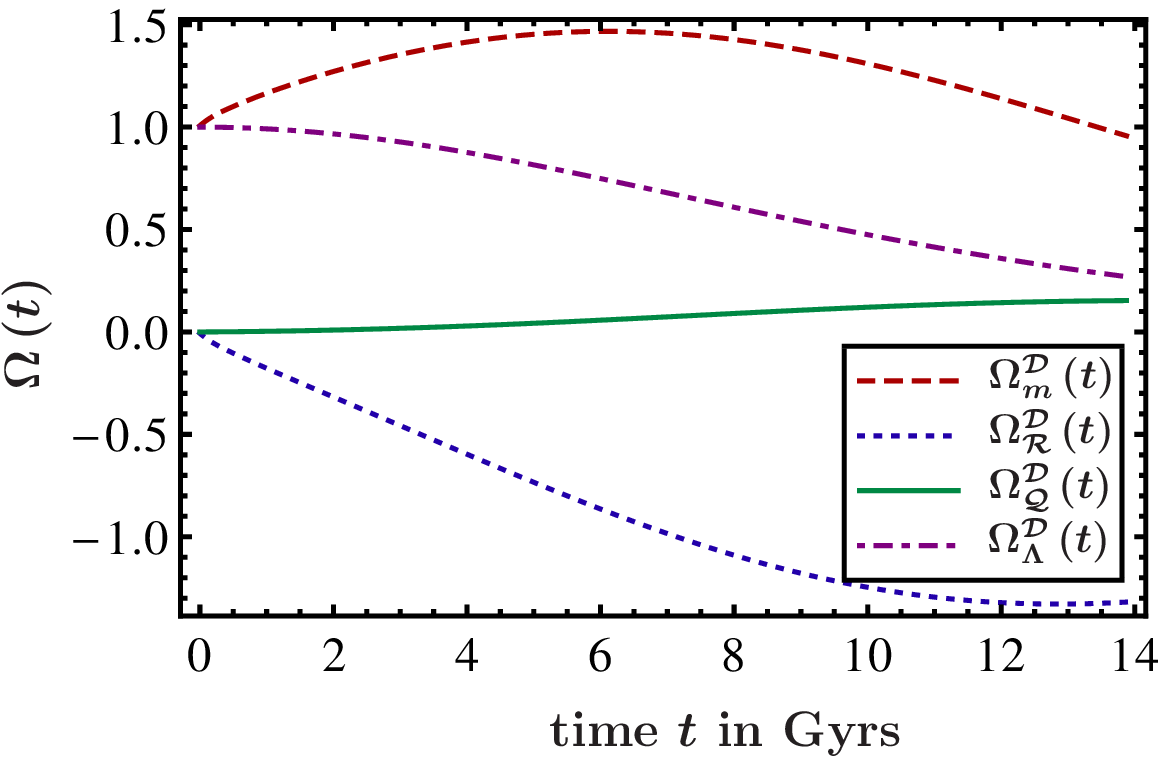}
    \caption{
    This figure corresponds to Fig.~\ref{fig:cospar50}, but shows the corresponding values for a collapsing domain of
    one--$\sigma$ fluctuations on the scale of $50$ Mpc. On this scale we appreciate a singular pancake collapse for the EdS model; Fig. \ref{fig:cospar25ratio} illustrates 
    that the backreaction term now becomes not only qualitatively but also quantitatively significant.
     \label{fig:cospar25}}
\end{figure*}
\begin{figure*}
  \includegraphics[width=0.47\textwidth]{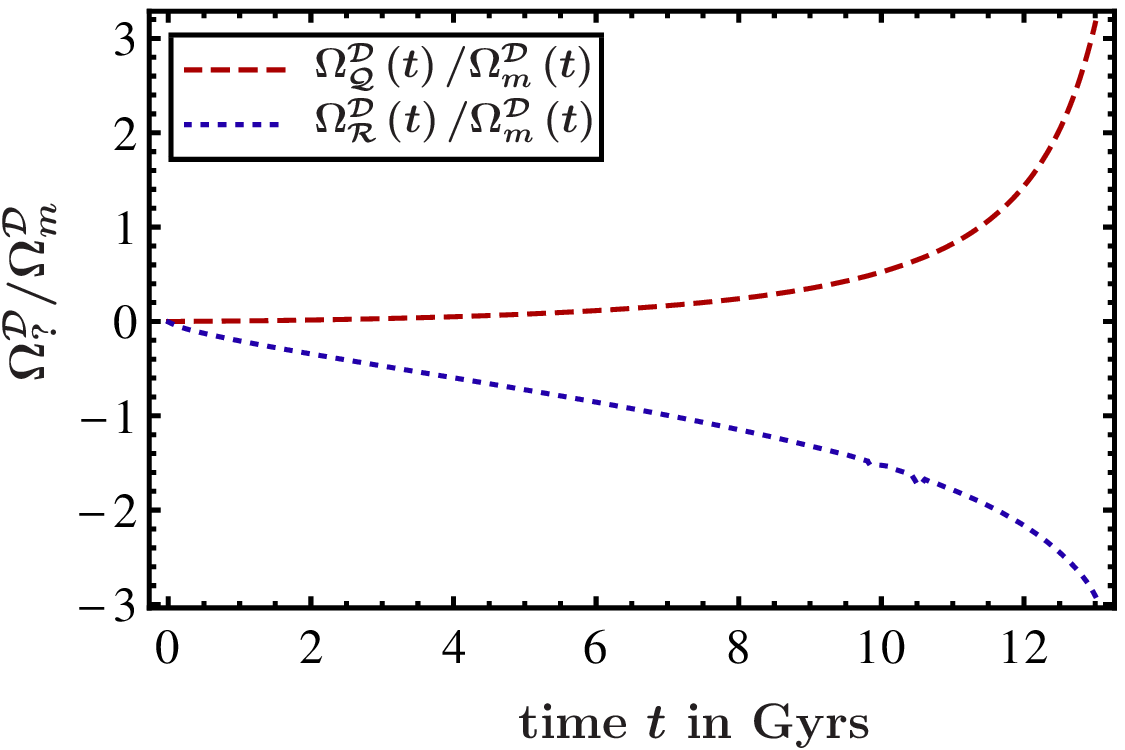}\hspace{20pt}
  \includegraphics[width=0.47\textwidth]{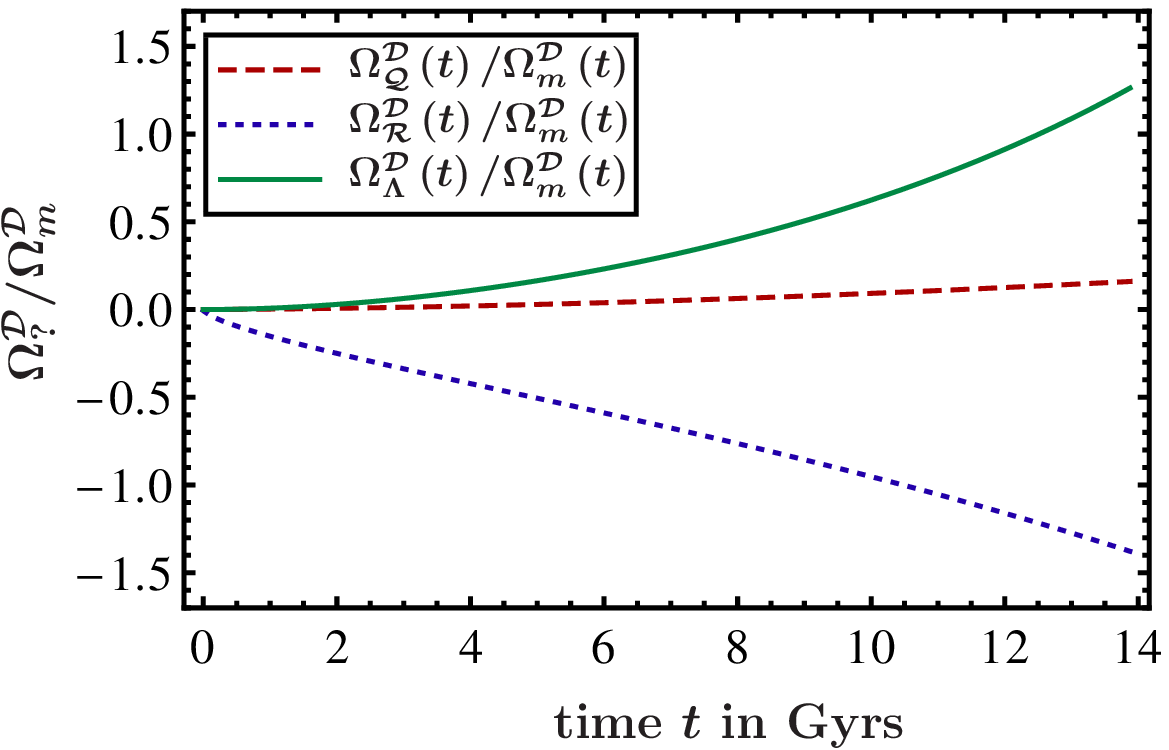}
    \caption{
    Ratio of the cosmic parameters for backreaction and curvature to the matter density parameter for a scale of $50$ Mpc as plotted in Fig.~\ref{fig:cospar25}. As during the pancake collapse $H_\cD$ becomes $0$, the parameters themselves are no longer well--defined. In the ratios plotted in this figure, however, $1/H_\cD^2$ cancels.
     \label{fig:cospar25ratio}}
\end{figure*}

Let us consider the cosmological density
parameters, Eq.~(\ref{eq:cospar}), describing the energy content balance of the
universe model.  In the average scenario they are domain dependent
quantities. In Fig.~\ref{fig:cospar100} we show their evolution with
cosmic time on an expanding typical (i.e. one--sigma) domain $200$ Mpc.
Fig.~\ref{fig:cospar100sig2} shows the situation again for an expanding domain of
$200$ Mpc, but here found with two--sigma probability.
In  Figs.~\ref{fig:cospar50} and \ref{fig:cospar25} we give examples of  smaller typical domains of $100$ and $50$ Mpc, 
but this time in the collapsing phase. 

For the case of the EdS background with
$\Omega_{\Lambda}=0=\Omega_{\Lambda}^{\cD}$ and $\Omega_{k}=0$, the
global matter density parameter is $\Omega_{m}=1$.  Regionally, however,
the plot shows that $\Omega_{m}^{\cD}$ is on $200$ Mpc domains typically
lower by $20$\%, i.e. this scale is dominated by underdense voids. 
In the relativistic framework the lower matter density is compensated by an emerging curvature parameter $\Omega_{\CR}^{\cD}$.

 It is interesting that the curvature deviation from the background curvature, $\CW_{\cD}$, a second--order quantity, can be 
  a lot bigger than the second--order quantity $\CQ_{\cD}$. This is due to the fact that
  in the perturbative expansion the second--order terms both scale as
  $a^{-1}$, and in view of the integrability condition
  Eq.~(\ref{eq:integrability-GR}), the second--order contribution
  to $\CW_\cD$ is a factor of $5$ larger than the second--order
  term of $\CQ_{\cD}$. Therefore, even a small backreaction
  contribution of only $2$\% will already lead to a $10$\%
  modification of the averaged curvature.

For the $\Lambda$CDM background the matter density parameter is also
reduced with respect to the background value and again we have
curvature emerging from a flat background. Today, however, it is not as
big as in the EdS case, since the cosmological constant dominates.

In both cases we see that, even though the backreaction contribution
stays tiny (as was already found in \cite{bks}), we have a considerable amount of curvature.
Comparing this curvature contribution with the flat geometry of the background on a given scale allows its 
interpretation in terms of `Dark Energy': the standard interpretation is that the matter distribution evolves on a flat
geometry; hence one would add a fundamental component to compensate the actually existing curvature that we
model here. In other words, the matter distribution has to be seen on a curved space section and not on a flat background.
Taking this point of view, the emerging curvature quantifies, on a given scale, the amount of `Dark Energy' that would be 
needed to compensate it in a quasi--Newtonian model. We find for the example of Fig.~\ref{fig:cospar100} that our model
predicts the existence of typical domains with a diameter of $200$ Mpc of about $1/4$ of the needed amount of `Dark Energy'. This amount can increase to 
$0,4$, if the $200$ Mpc domain is slightly untypical (i.e. found with a two--sigma probability, Fig.~\ref{fig:cospar100sig2}).

Figs.~\ref{fig:cospar50}, \ref{fig:cospar25} and \ref{fig:cospar25ratio} show that the backreaction can have the opposite effect by looking at smaller scales. A typical, collapsing domain produces
on average a positive curvature (corresponding to a negative curvature density parameter). In light of the interpretation above, the collapse produces
a large amount of `Dark Matter' in the form of positive curvature. While the `cosmological parameters' start to lose their sense on this scale, we can clearly see, e.g. in Fig.~\ref{fig:cospar50}, that
the curvature contribution is of the order of the density contribution, since it compensates the produced over--densities. The physical interpretation is also clear: over--densities are hosted in positive--curvature environments. The order of magnitude of `Dark Matter' is
also comparable with that of `Dark Energy' in expanding domains, i.e. we can also here say that a substantial fraction of the density parameters is contained in the curvature parameter, while it is of comparable magnitude and not dominating. 

\begin{figure*}
  \includegraphics[width=0.47\textwidth]{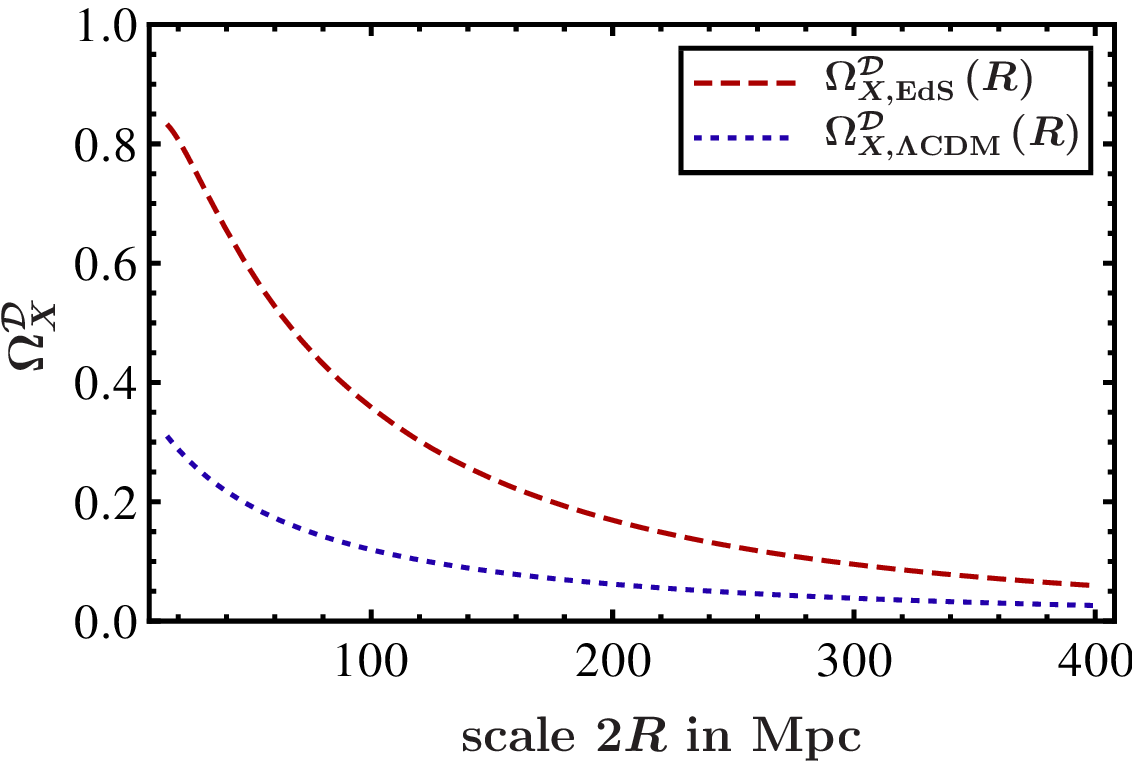}\hspace{20pt}
  \includegraphics[width=0.47\textwidth]{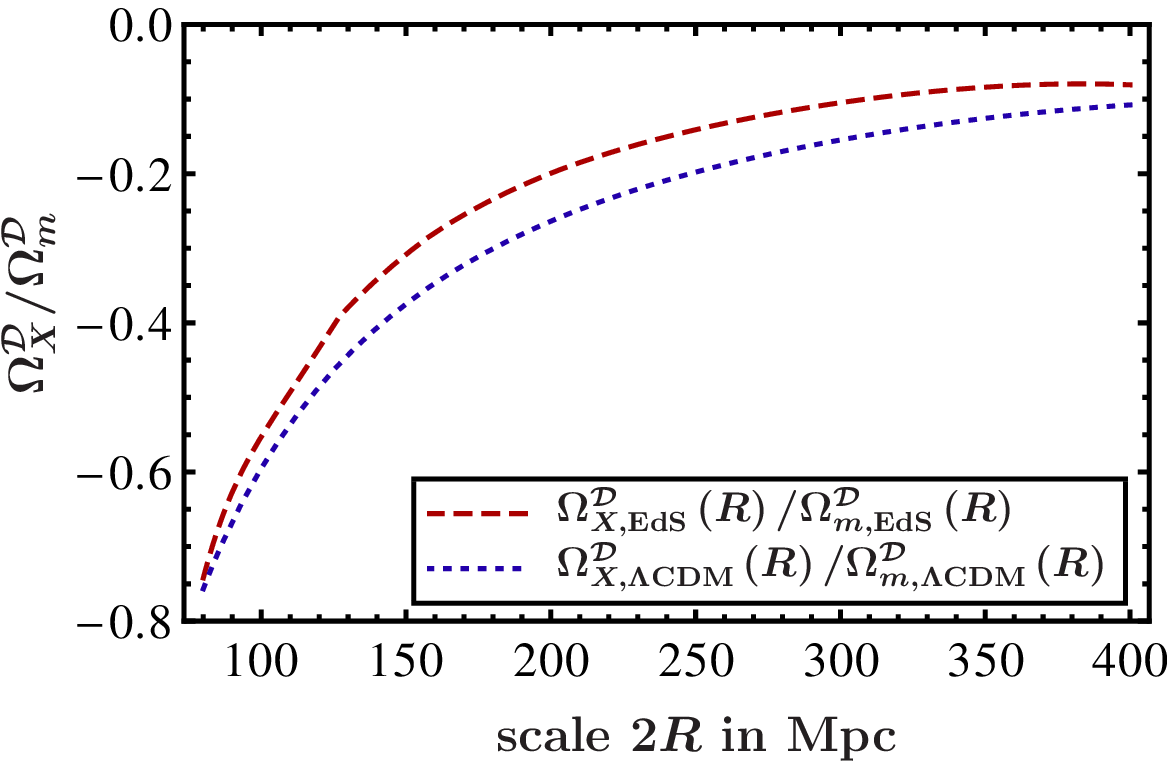}
    \caption{
    Scale--dependence of the $X-$Matter component, $\Omega_X^\cD := \Omega_\CQ^\cD + \Omega_\CR^\cD$, {\em today} for {\it typical} (one--sigma) domains in the averaged RZA. Left: underdense regions. Right: overdense regions. As in this case the Hubble parameter goes through zero for small scales, we plot the ratio of the $X-$Matter to the matter density parameter. In this ratio $H_\cD$ drops out.
     \label{fig:XMatScale}}
\end{figure*}

Given the conservative assumptions of our model these values point to a highly significant effect of backreaction.
Except in particular LTB models, where `Dark Energy' can be fully replaced, the only generic model that carries such a large effect
has been investigated by Enqvist {\em et al.} \cite{Enqvist:gradientexpansion} by employing gradient expansion techniques that allow
to go substantially beyond standard (Eulerian) perturbation methods \cite{salopek,cornelius}. 
These latter have to assume a universe model that stays close to the assumed background.
For comparison, we wish to point the reader to a figure in \cite{Enqvist:gradientexpansion}: Fig.~\ref{fig:XMatScale} corresponds to their Fig.~4 and shows the 
scale dependence of the $X-$Matter component $\Omega_X^\cD := \Omega_\CQ^\cD + \Omega_\CR^\cD$ today: the $X-$Matter produces `Dark Energy' on expanding, underdense domains (left panel), starts to compete with the matter density on scales below $100$ Mpc and it is still significant ($6$\%) on domains of $400$ Mpc; it produces `Dark Matter' on collapsing, overdense domains (right panel), 
starts to compete with the matter density again below the scales of $100$ Mpc, and shows similar significance as `Dark Energy' on domains with diameter $400$ Mpc.
Note that the large--scale value is astonishingly big, since we are looking at a scale that is considered to comply with homogeneity. 

\section{Results, Conclusions and Outlook}
\label{sect:discussion}

Based on a nonlinear extrapolation of a first--order relativistic Lagrangian perturbation scheme investigated in \cite{rza1}, we 
modeled the fluctuations in extrinsic curvature in the form of  the kinematical backreaction term
$\CQ_\cD$ for the description of the average properties of irrotational dust models.
We provided backreaction and intrinsic curvature expressions as nonlinear functionals of the first--order Lagrangian deformation and
used them as an input for the general framework for the average dynamics. The large--scale behavior of the backreaction variables, 
${\cal Q}_\cD$ and ${\cal W}_\cD$, at second order is identical to the leading mode
in the second--order perturbation theory (proportional to $a^{-1}$ for an EdS background, while the leading curvature contribution is proportional to $a^{-2}$ in conformity with the perturbative calculations of \cite{li:scale}).
We showed that the backreaction model contains in limiting cases the Newtonian approximation investigated in \cite{bks},
as well as special classes of exact GR solutions. We argued that this backreaction model is a powerful approximation due to the successes of the corresponding elements of this approximation in Newtonian cosmology (like the large--scale performance of the Newtonian Zel'dovich approximation in comparison with N--body simulations, e.g. \cite{melott1,melott2}), and also due to the above--mentioned property of interpolating between exact GR solutions with orthogonal symmetries.

We discussed how substantial results of \cite{bks} can be translated to the relativistic context, the new feature being an emerging intrinsic averaged
curvature. This translation is possible due to the fact that the Lagrangian description offered in \cite{rza1} features a clear--cut Newtonian limit by sending the coefficients of the coframes 
to their Euclidean counterparts, $\eta^a_{\;i} \rightarrow f^a_{\;|i}$, where the existence of the vector field $f^a$ in the Newtonian case describes the embedding into a global Euclidean vector space. The so--called {\it electric part of the Lagrange--Einstein system} \cite{rza1} is formally identical to the Newtonian equations in Lagrangian form, where in the GR case the nonintegrability of the coframes is responsible for the emerging curvature. 

We quantified the fluctuations of various cosmological parameters on different spatial domains for standard model backgrounds (CDM and $\Lambda$CDM power spectrum), and we now focus on these quantitative results.

\subsection{Results}

In order to discuss the quantitative results of our investigation, let us concentrate on the averaged scalar curvature $\langle\CR\rangle_\cD$.
The principal idea is to compare the energy balance conditions, encoded in cosmological parameters (i) in the standard model and (ii) in the averaged model.
The balance condition for (i) is entirely defined through the chosen background cosmology and furnished by the normalized Hamilton constraint. For the two chosen backgrounds we simply have
\begin{equation}
\Omega_m + \Omega_{\Lambda} \;=\;1\;,
\end{equation} 
where initially $\Omega_m$ dominates and remains equal to $1$ for the EdS background, while the importance of $\Lambda$ increases in the late stages in the background of the `concordance model'. This balance has to be compared with the (scale dependent) balance of the averaged cosmology furnished by the normalized averaged Hamilton constraint,
\begin{equation}
\Omega_{m}^{\cD} + \Omega_X^\cD \;=\;1\;,
\end{equation} 
where $\Omega_X^\cD = \Omega_\CQ^\cD + \Omega_\CR^\cD$ comprises the emerging backreaction and curvature due to structure formation. This comparison allows us to interpret the $X-$Matter energy density content as a candidate, on some scale, for the contribution by $\Omega_{\Lambda}$ that is added {\it ad hoc} in the standard model.
One subtlety of this interpretation is related to the scale dependence of the averaged balance condition and also to its evaluation on a chosen background cosmology that is already supposed to contain a substantial amount of  `Dark Matter' (for both backgrounds) and, additionally, `Dark Energy' (for the `concordance model'). We will now discuss the values obtained at the present day, and denote the domain by $\cD_0$.

We found, in accord with previous analyses that were e.g. summarized in the review papers \cite{ellisbuchert,ellisFOCUS,kolbFOCUS,buchert:focus,rasanenFOCUS,chrisreview,buchertrasanen}, that on large underdense domains of the order of $100-400$ Mpc, fluctuations feature a negative averaged curvature due to the fact that
on these scales the universe model is void--dominated. Note that the curvature does not individually obey a conservation law like the density; only a combination of averaged curvature and backreaction is conserved \cite{buchertcarfora:curvature}. 

Let us look, for example, at a scale of $200$ Mpc.
While the contribution by kinematical backreaction $\Omega_{\CQ}^{\cD_{0}}$ remains small in this situation, the curvature density parameter (that is positive for negative curvature), $\Omega_{\CR}^{\cD_{0}}$ is, for one--sigma initial fluctuations, of the order of (e.g. on the EdS background) $20$\% and compensates the lowered matter density parameter on this scale, $\Omega_{m}^{\cD_{0}} \cong 80 \%$.
This result, being physically plausible, nevertheless paints a completely different picture from the one advocated in standard cosmology, e.g. \cite{ishibashiwald} (for a thorough 
treatment of the quality of quasi--Newtonian approximations see \cite{greenwald}), claiming that 
intrinsic curvature is irrelevant down to the scales of neutron stars (see discussions and estimates in \cite{kolb:voids,estim,rasanen:model,chrisFOCUS,buchertrasanen, clifton}). Since this latter prejudice led to the inclusion of a large amount of `Dark Energy' into the background,
as modeled by a cosmological constant in the $\Lambda$CDM model, we are entitled to take the EdS background without `Dark Energy' and interpret the negative averaged curvature contribution as a replacement of the `Dark Energy' in the standard interpretation, where curvature is strictly zero on all scales. We have quantified this contribution as less than $20$\% for one--sigma fluctuations on $200$ Mpc, which
is -- in this model -- not enough to compensate the missing `Dark Energy' of the order of $73$\%. (Note that when speaking about `Dark Energy' we do not imply that the $X-$Matter component always produces an accelerating universe model.) For two--sigma fluctuations on this scale we find $35$\% of the needed amount, but due to the architecture of our model, as discussed below, this contribution falls off rapidly on very large scales. We may compare this regional behavior in the sense of so--called `void models'  that assume we are living in a large underdense region, see e.g. \cite{bolejkoFOCUS,krasinski.bolejko,bolejkoCUP} and references therein. In the spirit of these models our generic model `explains away' $1/2$ of the `Dark Energy' for a slightly untypical region (two--sigma) on the scale of $200$ Mpc. 

Before we explain why we cannot expect this behavior on larger scales in our model, we briefly sketch the
opposite behavior of the backreaction on smaller scales, where a strongly anisotropic collapse into sheet--like and filamentary--like structures features a negative backreaction term $\CQ_\cD$ and a positive curvature term $\average{\CR}$. This situation occurs e.g. on the scale of $50$ Mpc, below which the $X-$Matter component starts to compete with the matter density and mimics the presence of `Dark Matter'. This result points to the high relevance of the backreaction effect for the explanation of `Dark Matter' in 
over--densities, while this effect decays on large scales leaving only a few percent remnant on an assumed homogeneity scale of about $400$ Mpc. This latter is comparable with the `Dark Energy' remnant on such scales. Both effects thus are small but still alter the large--scale cosmological parameters at the percent level, see Fig.~\ref{fig:XMatScale}.
\subsection{Conclusions}

There are several serious shortcomings of our model that has a limited architecture as compared with the general situation, below discussed as {\it the background problem}, and also our model is, due to the choice of CDM and $\Lambda$CDM initial conditions, exploited in a regime where the backreaction effect is not yet effective on large scales, below discussed as {\it the amplitude problem}.

\subsubsection{The background problem}

In our model the background is "fictitious'': while our backreaction models have a generic structure, our quantitative interpretations largely depend on the choice of background
\cite{kolb:backgrounds}.
The reason is that both of our backgrounds contain a large amount of `Dark Matter', while the effect we study produces kinematical contributions that act in a similar way. 
The same argument holds for the background containing `Dark Energy'. 
The implementation of a physical background, defined through the spatial average of the considered fluctuations, is the next step to be envisaged (compare the first results obtained in \cite{roy:perturbations}). Such a physical background interacts with the backreaction models and thus changes their interpretation. While in the corresponding Newtonian model \cite{bks} the background forms the average distribution due to its torus architecture \cite{buchertehlers}, i.e. the scale factor $a_\cD \equiv a$ on the periodicity scale, the relativistic model studied here must be considered a hybrid construction featuring a volume--scale factor $a_\cD$ and a background scale factor $a$ that only approximately matches the volume--scale factor on the largest scales as a result of large--scale remnants of backreaction.
An indication for this shortcoming is also that the backreaction model, seen on the
FLRW background, only covers very restricted subcases of exact GR solutions, unlike
the situation in Newtonian cosmology, where the plane and spherical collapse is in general
covered by the backreaction model. 

\subsubsection{The amplitude problem}

Closely related to the first problem of a background containing a dominating `Dark Matter' component, we here point out a serious mismatch between 
the time of onset of the actual large--scale backreaction effect and the age of the Universe as it is determined by the background model.
We can take as an indicator the onset of an accelerating period that will set in as a result of dominance of backreaction effects over the density. 
Such an acceleration may not be needed to explain observational data; also, it is a strong requirement in view of the smallness of the kinematical backreaction term, which is the only backreaction component entering the volume acceleration equation, while the impact of $X$-matter is considerably stronger.
From our model we infer that this acceleration phase is seen in the future on scales roughly above $40$ Mpc. 
This result confirms the amplitude problem already
reported in \cite{multiscale}, if we suppose that backreaction were to replace `Dark Energy' completely.
Also, if we would consider the analyses presented in this paper in the period of onset of the effect, our quantitative conclusions on the amount of $X-$Matter would substantially change. Thus, all of our results remain conservative -- although visible in the fluctuation properties on intermediate and small scales -- since the actual effect is postponed to the future due to our standard initial data setting. 

We can understand this by a simple consideration: the averaged density decays as $a_\cD^{-3}$, the constant curvature (contained in the full averaged Ricci curvature) decays as $a_\cD^{-2}$, the leading large--scale mode of the backreaction model (the kinematical backreaction ${\cal Q}_\cD$) decays as 
$a_\cD^{-1}$ and, finally, the nondecaying cosmological constant. Thus, we expect several periods of onset of dominance of (i) the constant--curvature part and later (ii) the backreaction contribution, assuming that the cosmological constant is set to zero. As we saw, this latter period lies in the future for our initial data setting.

There are furthermore limitations due to the model's input of a first-order deformation. Obviously, this limits the range of applicability for 
the full backreaction regime that here lies in the future, but would anyway not be covered by our extrapolation of a first--order solution. 
Generally, our model is applicable only to large scales, well beyond the scales of virialized objects. The matter distribution is assumed to be fairly smooth and does not take into account the small-scale discretization. The importance of the modeling of small-scale structure has to be emphasized. For example, the determination of the actual volume fraction in devoid regions is needed to precisely quantify the contribution of a negative curvature (and in turn of the contribution to what would be interpreted as `Dark Energy' in the standard model), \cite{rasanen:peakmodel,multiscale,wiltshireFOCUS}, and this sensibly depends on the high-resolution modeling of small-scale structure. Collapse models, as the one based on \cite{bks}, \cite{kerscher:abundance}, have to be refined by e.g. including velocity dispersion. A corresponding argument, making use of the evolution of the fraction of virialized objects together with an attempt to also model the light cone effect, may give an indication of its quantitative importance \cite{boudDE}. 

\subsubsection{Outlook}

Clearly, in view of the remarks above, further major efforts are needed, and the present investigation can only be considered as a further step towards quantifying backreaction effects.
We expect, however, that the elements we provided in this work will prove useful for further considerations and the construction of improved models.
At any rate, a model based on the RZA is expected to realistically model the large--scale skeleton, and so lies at the heart of any model for structure formation including fully relativistic numerical simulations.
Furthermore, the above remark has to be seen in junction with another major effort to reinterpret observational data in the relativistic models: the subtleties and complexity of the interpretation of curvature energies as `Dark Energy' and `Dark Matter', seen on intermediate scales in our investigation, are not only a challenge in view of the scale dependence of the effect, but imply important changes of light propagation in these models \cite{morphon:obs,rasanen:lightpropagation,fleury}. Here again the precise modeling of small-scale structure is crucial.

The present investigation reveals, in a quantified way, how we have to proceed in order to master inhomogeneous universe models. We identify four key questions to be addressed: first, how can we deal with the `kinematical Dark Matter' produced in the relativistic models on all scales, and how does this alter the initial conditions and their compatibility with CMB constraints? Second, how can we improve the model to quantify the effect on smaller scales than those accessible here? Third, how can we implement a perturbation scheme on the physical background, given by the average of the model?
Fourth, how are observational data reinterpreted in the relativistic models?

The answer to the first three questions would alter the time scale of the background, of structure formation and of our interpretation of backreaction. Would this imply a shift of the backreaction regime into an epoch before today? Interpreting the onset of an acceleration period as coinciding with an apparent acceleration is highly problematic \cite{bolejkoandersson}, we are tempted to require such a shift to solve the `coincidence problem', but an acceleration of the model may after all not be needed to explain observational data. 
The answer to the third question will deliver backreaction models that interact with the background; they will contain, e.g., the volume scale factor (and not the FLRW scale factor), which itself depends on the backreaction terms. An iterative scheme results that would substantially modify the time scales of our results.
In turn, the amplitude of large--scale backreaction could substantially exceed the $6$\% effect already present in our model on an assumed homogeneity scale of $400$ Mpc.

Finally, let us emphasize again that we constructed our model by keeping basic cornerstones of the standard model in the early stages of the cosmic evolution. Insufficiency of late--time backreaction effects may indicate that a more drastic paradigmatic change is needed that takes into account backreaction effects in the Early Universe: consequences of emerging average properties from inhomogeneous inflationary models \cite{inflation}, and cosmological models that are built on global principles like globally stationary cosmologies \cite{buchert:static}, which inherently explain a large backreaction component contained in the physical background, have thus far not been thoroughly analyzed.


\subsection*{Acknowledgements}

{\small The work of TB was supported by "F\'ed\'eration de Physique Andr\'e--Marie Amp\`ere, Lyon'', 
and was conducted within the "Lyon Institute of Origins'' under Grant No. ANR-10-LABX-66.
CN acknowledges support by the {\'E}cole Doctorale Lyon.
AW was partially supported by the DFG under Grant No. GRK 881, and he acknowledges support from CRALyon during working visits.
We thank Martin Kerscher for providing his Newtonian code. We thank him and also Cornelius Rampf, Boudewijn Roukema and Xavier Roy for valuable comments on the manuscript.
}{\small \par}

\appendix

\section{Demonstration of Propositions 1 and 2}

\label{sect:AppendixProp} \textbf{Proposition 1.} 
\begin{equation}
  ^{{\rm RZA}}{\cal Q_{D}}=0\Leftrightarrow\left\{ \begin{array}{ll}
      \caverage{\initial{\inII}}={\displaystyle \frac{1}{3}\caverage{\initial{\inI}}^{2}\;,}\\
      \\\caverage{\initial{\inIII}}={\displaystyle
        \frac{1}{27}\caverage{\initial{\inI}}^{3}\;.}\end{array}\right.
\label{eq:propos1}
\end{equation}

\textit{Proof:}\\
\textbullet{}\ With Eq.~(\ref{resultQ2}), the demonstration
of the vanishing of the backreaction term, given the right--hand--side, is straighforward.\\
\textbullet{}\ For given vanishing backreaction, i.e. supposing that: $\forall t,\ ^{{\rm
    RZA}}{\cal Q_{D}}=0$ we proceed as follows.  Excluding the case $\dot{\xi}(t)=0$
$\forall t$, using Eq.~(\ref{resultQ2}), the vanishing of the
backreaction term is equivalent to: 
\begin{equation} \forall t,\ \ \
  \gamma_{1}+\xi(t)\gamma_{2}+\xi^{2}(t)\gamma_{3}=0\;.
\label{eq:dem-propos1-begin}
\end{equation}
As the monomials in $\xi$ are linearly independent, already the
coefficients have to vanish. Therefore,
$\gamma_{1}=\gamma_{2}=\gamma_{3}=0$, what, by the definition of
Eq.~(\ref{eq:GammaDef}), is equivalent
to the right--hand--side of Eq.~(\ref{eq:propos1}). \Marque{\SymPreuv}\\

\textbf{Proposition 2.} 
\begin{equation} ^{{\rm RZA}}{\cal
    Q_{D}}=0\Leftrightarrow\left\{ \begin{array}{ll}
      \average{^{\rm RZA}{\rm II}\left(\Theta_{\ j}^{i}\right)}={\displaystyle \frac{1}{3}\average{^{\rm RZA}{\rm I}\left(\Theta_{\ j}^{i}\right)}^{2}\;,}\\
      \\\average{^{\rm RZA}{\rm III}\left(\Theta_{\
            j}^{i}\right)}={\displaystyle
        \frac{1}{27}\average{^{\rm RZA}{\rm I}\left(\Theta_{\
              j}^{i}\right)}^{3}\;.}\end{array}\right.
\label{eq:propos2}
\end{equation}

\textit{Proof:}\\
\textbullet{}\ Once again the vanishing of the backreaction term, given the right--hand--side, is straighforward.\\
\textbullet{}\ For the other way we use (Prop.~\ref{propos1}):
if, $\forall t,\ ^{{\rm RZA}}{\cal Q_{D}}=0$, then one can write
$\caverage{\initial{\inII}}$ and $\caverage{\initial{\inIII}}$ as
functions of $\caverage{\initial{\inI}}$; with
Eqs.~(\ref{eq:inIII-pelicular_inIII}) and (\ref{eq:RZA-peculiar-Inv})
we have: %
\begin{eqnarray}
  \caverage{{\frak J}} & = & \left(1+\frac{\xi}{3}\caverage{\initial{\inI}}\right)^{3}\;,\nonumber \\
  \average{^{\rm RZA}{\rm I}\left(\Theta_{\ j}^{i}\right)} & = & 3\frac{\dot{a}}{a}+{\cal X}\;,\nonumber \\
  \average{^{\rm RZA}{\rm II}\left(\Theta_{\ j}^{i}\right)} & = & 3\left(\frac{\dot{a}}{a}\right)^{2}+2\frac{\dot{a}}{a}{\cal X}+\frac{1}{3}{\cal X}^{2}\;,\nonumber \\
  \average{^{\rm RZA}{\rm III}\left(\Theta_{\ j}^{i}\right)} & = & \left(\frac{\dot{a}}{a}\right)^{3}+\left(\frac{\dot{a}}{a}\right)^{2}{\cal X}+\frac{1}{3}\frac{\dot{a}}{a}{\cal X}^{2}+\frac{1}{27}{\cal X}^{3}\;,\nonumber \\
\end{eqnarray}
where ${\cal X}=\dot{\xi}\caverage{\initial{\inI}} / (1+{\displaystyle
      \frac{\xi}{3}\caverage{\initial{\inI}})}$.
This finally leads to the equalities of the right--hand--side of
Eq.~(\ref{eq:propos2}).  \Marque{\SymPreuv}

\section{Orthonormal basis representation}
\label{sec:Appendix-B}

We will, in this Appendix, give all the relevant expressions for the case where 
we employ the standard assumption of orthonormal frames (Option 1 in the text, see \ref{sec:RZA}).
The coframes are in this case given by
\begin{equation}
^{{\rm RZA}}\eta_{\ i}^{a}\left(t,X^{k}\right):=a\left(t\right)\left(\delta_{\; i}^{a}+P{^{a}}_{i}+\xi(t)\dot{P}_{\ i}^{a}\right)\;,
\end{equation}
which combine, via $\delta_{ab}$, to the complete metric coefficients 
\begin{equation}
g_{ij}:=\delta_{ab}\,\eta_{\; i}^{a}\eta_{\; j}^{b}\;.
\end{equation}
We evaluate for this case some relevant fields furnishing the RZA and its average properties:
\begin{itemize}
\item the coefficients of the metric tensor: 
\begin{eqnarray}
    ^{{\rm RZA}}g_{ij}(t,X^{k}) & = & a^{2}(t)\left\{ \delta_{ij}+2P_{(ij)}+P_{ki}P{^{k}}_{j}\right.\nonumber \\
    &  & \left.+2\xi(t)\left(\dot{P}_{(ij)}+{P}_{k(i}\dot{P}_{\; j)}^{k}\right)\right.\nonumber \\
    & & \left.+\xi^{2}(t)\dot{P}_{ki}\dot{P}_{\;
        j}^{k}\right\} \ .
\label{eq:RZA-metric}
\end{eqnarray}
\item the local volume deformation: 
\begin{eqnarray}
    ^{{\rm RZA}}{\rm J} & := & \frac{1}{6}\epsilon_{abc}\epsilon^{ijk}\eta_{\ i}^{a}\eta_{\ j}^{b}\eta_{\ k}^{c}\nonumber \\
    & = & a^{3}(t){\frak J}\ .
\label{eq:RZA-jacobian}
\end{eqnarray}
\item the first scalar invariant of the expansion
  tensor: 
\begin{eqnarray}
    ^{{\rm RZA}}{\rm I}(\Theta_{\ j}^{i}) & := & \frac{1}{2{\rm J}}\epsilon_{abc}\epsilon^{ijk}\dot{\eta}_{\ i}^{a}\eta_{\ j}^{b}\eta_{\ k}^{c}\nonumber \\
    & = & 3\frac{\dot{a}(t)}{a(t)}+\frac{\dot{{\frak J}}}{{\frak J}}\
    .
\label{eq:RZA-I}
\end{eqnarray}
\item the second scalar invariant of the expansion
  tensor: 
\begin{eqnarray}
    ^{{\rm RZA}}{\rm II}(\Theta_{\ j}^{i}) & := & \frac{1}{2{\rm J}}\epsilon_{abc}\epsilon^{ijk}\dot{\eta}_{\ i}^{a}\dot{\eta}_{\ j}^{b}\eta_{\ k}^{c}\nonumber \\
    & = & 3\left(\frac{\dot{a}(t)}{a(t)}\right)^{2}+2\frac{\dot{a}(t)}{a(t)}\frac{\dot{{\frak J}}}{{\frak J}}\nonumber \\
    & + &
    \frac{1}{2}\left(\frac{\ddot{{\frak J}}}{{\frak J}}-\frac{\ddot{\xi}(t)}{\dot{\xi}(t)}\frac{\dot{{\frak J}}}{{\frak J}}\right)\
    .
\label{eq:RZA-II}
\end{eqnarray}
\item the third scalar invariant of the expansion
  tensor: 
\begin{eqnarray}
    ^{{\rm RZA}}{\rm III}(\Theta_{\ j}^{i}) & := & \frac{1}{6{\rm J}}\epsilon_{abc}\epsilon^{ijk}\dot{\eta}_{\ i}^{a}\dot{\eta}_{\ j}^{b}\dot{\eta}_{\ k}^{c}\nonumber \\
    & = & \left(\frac{\dot{a}(t)}{a(t)}\right)^{3}+\left(\frac{\dot{a}(t)}{a(t)}\right)^{2}\frac{\dot{{\frak J}}}{{\frak J}}
\label{eq:RZA-III}\\
    & + &
    \frac{1}{2}\frac{\dot{a}(t)}{a(t)}\left(\frac{\ddot{{\frak J}}}{{\frak J}}-\frac{\ddot{\xi}(t)}{\dot{\xi}(t)}\frac{\dot{{\frak J}}}{{\frak J}}\right)+\frac{\dot{\xi}^{3}(t)D_{{\rm
          ddd}}}{{\frak J}}\ .\nonumber 
\end{eqnarray}
\end{itemize}
We have introduced the functions: \[
{\frak J}(\xi(t),X^{i}):=S_{{\rm {\bf i}}}+\xi D_{{\rm d}}+\xi^{2}D_{{\rm
    dd}}+\xi^{3}D_{{\rm ddd}}\;,\]
\[
S_{{\rm {\bf i}}}:=S(\initial t,X^{k}):=1+{\rm I}_{{\rm {\bf i}}}+{\rm
  II}_{{\rm {\bf i}}}+{\rm III}_{{\rm {\bf i}}}\;,\]
\begin{eqnarray*}
  D_{{\rm d}}:=D_{{\rm d}}(\initial t,X^{k}) & := & {\rm I_{d}}\left(1+{\rm I}_{{\rm {\bf i}}}+{\rm II}_{{\rm {\bf i}}}\right)+{P}{^{i}}_{j}{P}{^{j}}_{k}\dot{P}_{\ i}^{k}\\
  && -\left(1+{\rm I}_{{\rm {\bf i}}}\right){P}{^{a}}_{b}\dot{P}_{\ a}^{b}\;,\\
  D_{{\rm dd}}:=D_{{\rm dd}}(\initial t,X^{k}) & := & {\rm II_{d}}\left(1+{\rm I}_{{\rm {\bf i}}}\right)+{P}{^{i}}_{j}\dot{P}_{\ k}^{j}\dot{P}_{\ i}^{k}\\
  && -\,{\rm I_{d}}{P}{^{a}}_{b}\dot{P}_{\ a}^{b}\;,\\
D_{{\rm ddd}}:=D_{{\rm ddd}}(\initial t,X^{k})&:=&{\rm III_{d}}\;,
\end{eqnarray*}
\begin{eqnarray*}
{\rm I}_{{\rm {\bf i}}}:={\rm I}({P}_{\ i}^{a})&:=&\frac{1}{2}\epsilon_{abc}\epsilon^{ijk}{
  P}{^{a}}_{i}\delta_{\ j}^{b}\delta_{\ k}^{c}\;,\\
{\rm II}_{{\rm {\bf i}}}:={\rm II}({P}_{\
  i}^{a})&:=&\frac{1}{2}\epsilon_{abc}\epsilon^{ijk}{
  P}{^{a}}_{i}{P}{^{b}}_{j}\delta_{\ k}^{c}\;,\\
{\rm III}_{{\rm {\bf i}}}:={\rm III}({P}_{\
  i}^{a})&:=&\frac{1}{6}\epsilon_{abc}\epsilon^{ijk}{
  P}{^{a}}_{i}{P}{^{b}}_{j}{P}{^{c}}_{k}\;,
\end{eqnarray*}
\begin{eqnarray*}
{\rm I_{d}}:={\rm I}(\dot{P}_{\
  i}^{a})&:=&\frac{1}{2}\epsilon_{abc}\epsilon^{ijk}\dot{P}_{\
  i}^{a}\delta_{\ j}^{b}\delta_{\ k}^{c}\;,\\
{\rm II_{d}}:={\rm II}(\dot{P}_{\
  i}^{a})&:=&\frac{1}{2}\epsilon_{abc}\epsilon^{ijk}\dot{P}_{\
  i}^{a}\dot{P}_{\ j}^{b}\delta_{\ k}^{c}\;,\\
{\rm III_{d}}:={\rm III}(\dot{P}_{\
    i}^{a})&:=&\frac{1}{6}\epsilon_{abc}\epsilon^{ijk}\dot{
    P}_{\ i}^{a}\dot{P}_{\ j}^{b}\dot{P}_{\
    k}^{c}\;.\\ 
\label{eq:RZA-function}
\end{eqnarray*}

\medskip\medskip
\noindent
Using Eq.~(\ref{eq:inIII-pelicular_inIII}) one can express the scalar
invariants of the peculiar--expansion tensor as a function of ${\frak J}$
and $\xi$: 
\begin{eqnarray} ^{{\rm RZA}}{\rm I}(\theta_{\
    j}^{i})=\frac{\dot{{\frak J}}}{{\frak J}}\;, \
    ^{{\rm RZA}}{\rm II}(\theta_{\ j}^{i})=
      \frac{1}{2}\left(\frac{\ddot{{\frak J}}}{{\frak J}}-\frac{\ddot{\xi}(t)}{\dot{\xi}(t)}\frac{\dot{{\frak J}}}{{\frak J}}\right),
\label{eq:RZA-pelicular_in}
\end{eqnarray}

\begin{equation}
^{{\rm RZA}}{\rm III}(\theta_{\
  j}^{i})=\frac{1}{6}\left(\frac{\dddot{{\frak J}}}{{\frak J}}-\frac{\dddot{\xi}(t)}{\dot{\xi}(t)}\frac{\dot{{\frak J}}}{{\frak J}}\right)-\frac{1}{2}\frac{\ddot{\xi}(t)}{\dot{\xi}(t)}\left(\frac{\ddot{{\frak J}}}{{\frak J}}-\frac{\ddot{\xi}(t)}{\dot{\xi}(t)}\frac{\dot{{\frak J}}}{{\frak J}}\right).
\end{equation}
Inserting Eqs. (\ref{eq:RZA-pelicular_in}) into
Eq. (\ref{eq:backreaction--card}), the backreaction term in the
`Relativistic Zel'dovich Approximation' reads:
\begin{equation} ^{{\rm
      RZA}}\CQ_{{\cal
      D}}=\frac{\caverage{\ddot{{\frak J}}}}{\caverage{{\frak J}}}-\frac{\ddot{\xi}}{\dot{\xi}}\frac{\caverage{\dot{{\frak J}}}}{\caverage{{\frak J}}}-\frac{2}{3}\left(\frac{\caverage{\dot{{\frak J}}}}{\caverage{{\frak J}}}\right)^{2},
\label{eq:backreaction-RZA}
\end{equation}
or finally, using the first equation of Eqs. (\ref{eq:RZA-function}):
\begin{eqnarray}
  & ^{{\rm RZA}}\CQ_{\cD}\;=\nonumber \\
  & {\displaystyle \frac{\dot{\xi}^{2}\left(\gamma_{1}+\xi\gamma_{2}+\xi^{2}\gamma_{3}\right)}{\left(\caverage{ S_{{\rm {\bf i}}}}+\xi\caverage{ D_{{\rm d}}}+\xi^{2}\caverage{ D_{{\rm dd}}}+\xi^{3}\caverage{ D_{{\rm ddd}}}\right)^{2}}\ ,}\nonumber
\label{resultQ1}
\end{eqnarray}
with: \begin{align}
  \begin{cases}
    \gamma_{1}:=2\caverage{ S_{{\rm {\bf i}}}}\caverage{D_{\rmdd}}-\frac{2}{3}\caverage{D_{\rmd}}^{2},\\
    \gamma_{2}:=6\caverage{ S_{{\rm {\bf i}}}}\caverage{D_{\rmddd}}-\frac{2}{3}\caverage{D_{\rmd}}\caverage{D_{\rmdd}},\\
    \gamma_{3}:=2\caverage{D_{\rmd}}\caverage{D_{\rmddd}}-\frac{2}{3}\caverage{D_{\rmdd}}^{2}.
    \end{cases}\end{align}

\quad


\begin{thebibliography}{2013}
 
\bibitem[L1]{rza1} T. Buchert and M. Ostermann: Lagrangian theory of structure formation in relativistic cosmology I: Lagrangian framework and definition of a nonperturbative approximation.  \emph{Phys. Rev. D} \textbf{86}, 023520 (2012).

\bibitem[BKS]{bks} T. Buchert, M. Kerscher and C. Sicka: Backreaction
of inhomogeneities on the expansion: The evolution of cosmological
parameters. \emph{Phys. Rev. D} \textbf{62}, 043525 (2000). 

\bibitem{bahcall:triangle} N. Bahcall, J.P. Ostriker, S. Perlmutter
  and P.J. Steinhardt: The cosmic triangle. \emph{Science} \textbf{284}, 1481 (1999). 

\bibitem{barrowgoetz}
J.D. Barrow and G. G\"otz: Newtonian no--hair theorems. 
\emph{Class. Quant. Grav.} \textbf{6}, 1253 (1989).

\bibitem{bildhauer:solutions}
S. Bildhauer, T. Buchert, and M. Kasai: Solutions in Newtonian cosmology -- the pancake theory with cosmological constant.
\emph{Astron. Astrophys.} \textbf{263},  23  (1992).

\bibitem{bolejko:szekeres} K. Bolejko: Volume averaging in the quasispherical Szekeres model.
\emph{Gen. Rel. Grav.} \textbf{41}, 1585 (2009). 

\bibitem{bolejkoandersson} K. Bolejko and L. Andersson: Apparent
and average acceleration of the Universe. \emph{J.C.A.P.} \textbf{10}, 003 (2008).

\bibitem{bolejkoCUP}
K. Bolejko, A. Krasi\'nski, C. Hellaby and M.--N. C\'el\'erier: Structures in the Universe by exact methods.
\emph{Cambridge Univ. Press} (2009).

\bibitem{bolejkoFOCUS}
K. Bolejko, M.--N. C{\'e}l{\'e}rier and A. Krasi\'nski: 
Inhomogenous cosmological models: exact solutions and their applications. 
\emph{Class. Quant. Grav.} \textbf{28}, 164002 (2011).

\bibitem{buchert:class} T. Buchert:  A class of solutions in Newtonian cosmology and the pancake theory.
\emph{Astron. Astrophys.} \textbf{223}, 9 (1989). 

\bibitem{buchertehlers} T. Buchert and J. Ehlers: Averaging inhomogeneous Newtonian cosmologies.
\emph{Astron. Astrophys.} \textbf{320},  1  (1997). 

\bibitem{buchert:onaverage} T. Buchert: On average properties of inhomogeneous cosmologies.
In: 9th JGRG Meeting,
Hiroshima 1999, Y. Eriguchi et al. (eds.), Hiroshima Univ. Press \emph{J.G.R.G.} {\bf 9}, 306--321 (2000),
arXiv:gr--qc/0001056  

\bibitem{buchert:onaverage1} T. Buchert: On average properties of inhomogeneous
fluids in general relativity:  dust cosmologies. \emph{Gen. Rel. Grav.}
\textbf{32}, 105 (2000). 

\bibitem{buchert:onaverage2} T. Buchert: On average properties of inhomogeneous
fluids in general relativity:  perfect fluid cosmologies. \emph{Gen. Rel.
Grav.} \textbf{33}, 1381 (2001).

\bibitem{buchert:static} T. Buchert: On globally static and stationary
cosmologies with or without a cosmological constant and the Dark Energy
problem. \emph{Class. Quant. Grav.} \textbf{23}, 817 (2006). 

\bibitem{buchert:darkenergy} T. Buchert: Dark Energy from structure --
a status report. \emph{Gen. Rel. Grav.} \textbf{40}, 467 (2008). 

\bibitem{buchert:focus} T. Buchert: Toward physical cosmology: focus on inhomogeneous geometry and its nonperturbative effects. 
\emph{Class. Quant. Grav.} \textbf{28}, 164007 (2011).

\bibitem{buchertcarfora:curvature} T. Buchert and M. Carfora: On the curvature
of the present--day Universe. \emph{Class. Quant. Grav.} \textbf{25}, 195001
(2008). 

\bibitem{buchertgoetz}
T. Buchert and G. G\"otz: A class of solutions for self--gravitating dust in Newtonian gravity. 
\emph{J. Math. Phys.} \textbf{28}, 2714 (1987).

\bibitem{estim} T. Buchert, G.F.R. Ellis and H. van Elst: Geometrical
order--of--magnitude estimates for spatial curvature in realistic
models of the Universe. \emph{Gen. Rel. Grav.} \textbf{41}, 2017 (2009). 

\bibitem{morphon} T. Buchert, J. Larena and J.--M. Alimi: Correspondence
between kinematical backreaction and scalar field cosmologies -- the
'morphon field'. \emph{Class. Quant. Grav.} \textbf{23}, 6379 (2006). 

\bibitem{melott1} T. Buchert, A.L. Melott and A.G. Wei{\ss}: Testing higher--order Lagrangian perturbation theory against numerical simulations I. Pancake models.
\emph{Astron. Astrophys.} \textbf{288}, 349 (1994).

\bibitem{inflation} T. Buchert and N. Obadia: Effective inhomogeneous inflation: curvature inhomogeneities of the Einstein vacuum.
\emph{Class. Quant. Grav.} \textbf{28}, 162002 (2011).

\bibitem{buchertrasanen}
T. Buchert and S. R\"as\"anen: Backreaction in late--time cosmology.
\emph{Ann. Rev. Nucl. Part. Sci.} \textbf{62}, 57 (2012).

\bibitem{chandra:blackholes} S. Chandrasekhar: The Mathematical Theory
  of Black Holes, Clarendon Press Oxford (1992).

\bibitem{clarkson}
C. Clarkson, K. Ananda and J. Larena: The influence of structure formation on the cosmic expansion.
\emph{Phys. Rev. D} \textbf{80}, 083525 (2009).

\bibitem{chrisreview} C. Clarkson, G.F.R. Ellis, J. Larena and O. Umeh: 
Does the growth of structure affect our dynamical models of the Universe? The averaging, backreaction, and fitting problems in cosmology.
\emph{Rep. Prog. Phys.} \textbf{74}, 112901 (2011).

\bibitem{chrisFOCUS} C. Clarkson and O. Umeh: Is backreaction really small within concordance cosmology? 
\emph{Class. Quant. Grav.} {\bf 28}, 164010 (2011).

\bibitem{clifton} 
T. Clifton: Backreaction in relativistic cosmology.
\emph{Int. J. Mod. Phys. D} \textbf{22}, 133004 (2013). 

\bibitem{ellisFOCUS} 
G.F.R. Ellis: Inhomogeneity effects in cosmology.
\emph{Class. Quant. Grav.} \textbf{28}, 164001 (2011).

\bibitem{ellisbuchert} G.F.R. Ellis and T. Buchert: The Universe
seen at different scales. \emph{Phys. Lett. A.} (Einstein Special Issue)
\textbf{347}, 38 (2005). 

\bibitem{Enqvist} K. Enqvist: Lema\^\i tre Tolman Bondi model and
accelerating expansion. \emph{Gen. Rel. Grav.} \textbf{40}, 451 (2008). 

\bibitem{Enqvist:gradientexpansion}
K. Enqvist, S. Hotchkiss and G. Rigopoulos: A gradient expansion for cosmological backreaction.
\emph{J.C.A.P.} \textbf{03}, 026 (2012).

\bibitem{fleury}
P. Fleury, H. Dupuy and J.--P. Uzan: Interpretation of the Hubble diagram in a non--homogeneous universe.
\emph{arXiv:1302.5308} (2013).

\bibitem{ellis:birkhoff1}
R. Goswami and G.F.R. Ellis: Almost--Birkhoff theorem in general relativity.
\emph{Gen. Rel. Grav.} \textbf{43}, 2157 (2011). 

\bibitem{ellis:birkhoff2}
R. Goswami and G.F.R. Ellis: Birkhoff theorem and matter.
\emph{Gen. Rel. Grav.} \textbf{44}, 2037 (2012). 

\bibitem{greenwald}
S.R. Green and R.M. Wald: Newtonian and relativistic cosmologies.
\emph{Phys. Rev. D.} \textbf{85}, 063512 (2012).

\bibitem{szekeres:obs1}
M. Ishak, J. Richardson, D. Garred, D. Whittington, A. Nwankwo and R.A. Sussman: 
Dark Energy or apparent acceleration due to a relativistic cosmological model more complex than FLRW ?
\emph{Phys. Rev. D} \textbf{78}, 123531 (2008).

\bibitem{szekeres:obs3}
M. Ishak and A. Peel: The growth of structure in the Szekeres Class--II inhomogeneous cosmological models and the matter--dominated era.
\emph{Phys. Rev. D} \textbf{85}, 083502 (2012).

\bibitem{ishibashiwald} A. Ishibashi and R.M. Wald: Can the acceleration of our universe be explained by the effects of inhomogeneities? 
\emph{Class. Quant. Grav.} \textbf{23}, 235 (2006). 

\bibitem{kasai95}
M. Kasai: Tetrad--based perturbative approach to inhomogeneous universes: a general relativistic version of the Zel'dovich approximation.
\emph{Phys. Rev. D} \textbf{52}, 5605 (1995).

\bibitem{kerscher:abundance}
Kerscher, M., Buchert, T., Futamase, T.:
On the abundance of collapsed objects.
\emph{Astrophys. J.} {\bf 558}, L79 (2001). 

\bibitem{kolb:voids} E.W. Kolb, V. Marra and S. Matarrese: Description
of our cosmological spacetime as a perturbed conformal Newtonian metric
and implications for the backreaction proposal for the accelerating
Universe. \emph{Phys. Rev. D} \textbf{78} 103002 (2008). 

\bibitem{kolb:backgrounds} E.W. Kolb, V. Marra and S. Matarrese:
Cosmological background solutions and cosmological backreactions.
\emph{Gen. Rel. Grav.} \textbf{42} 1399 (2010). 

\bibitem{kolbFOCUS} 
E.W. Kolb: Backreaction of inhomogeneities can mimic dark energy.
\emph{Class. Quant. Grav.} \textbf{28}, 164009 (2011).

\bibitem{krasinski:szekeres} 
A. Krasi\'nski: Geometry and topology of the quasi--plane Szekeres model.
\emph{Phys. Rev. D} \textbf{78}, 064038 (2008); erratum \emph{Phys.Rev. D} \textbf{85}, 069903(E) (2012).

\bibitem{krasinski.bolejko}
A. Krasi\'nski and K. Bolejko: Exact inhomogeneous models and the drift of light rays induced by non--symmetric flow of the cosmic medium.
\emph{arXiv:1212.4697} (2012).

\bibitem{morphon:obs} J. Larena, J.--M. Alimi, T. Buchert, M. Kunz
and P.--S. Corasaniti: Testing backreaction effects with observations.
\emph{Phys. Rev. D} \textbf{79}, 083011 (2009). 

\bibitem{lemaitre-tolman-bondi} G. Lema\^{i}tre, Annales Soc. Sci.
  Brux. Ser. I Sci. Math. Astron. Phys. A 53:51 (1933) (in French)\\
  G. Lema\^{i}tre, Gen. Rel. and Grav., 29, 5 (1997) (reprint).\\
  R.C. Tolman, Proc. Nat. Acad. Sci. 20:169 (1934).\\
  H. Bondi, Mon. Not. Roy. Astron. Soc. 107: 410 (1947).

\bibitem{li:scale} N. Li and D.J. Schwarz: Scale dependence of
  cosmological backreaction. \emph{Phys. Rev. D} \textbf{78}, 083531 (2008). 

\bibitem{matarrese&terranova} 
S. Matarrese and D. Terranova: Post--Newtonian cosmological dynamics in Lagrangian coordinates.
\emph{Mon. Not. Roy. Astron. Soc.} \textbf{283}, 400 (1996). 

\bibitem{melott2} A.L. Melott, T. Buchert and A.G. Wei{\ss}: Testing higher--order Lagrangian perturbation theory against numerical simulations II. Hierarchical models.
\emph{Astron. Astrophys.} \textbf{294}, 345 (1995).

\bibitem{meures.bruni} N. Meures and M. Bruni: 
Exact nonlinear inhomogeneities in $\Lambda$CDM cosmology.
\emph{Phys. Rev. D} \textbf{83}, 123519 (2011).

\bibitem{RZAandLTB}
M. Morita, K. Nakamura and M. Kasai: Relativistic Zel'dovich approximation in a spherically symmetric model.
\emph{Phys. Rev. D} \textbf{57}, 6094 (1998); erratum \emph{Phys. Rev. D} \textbf{58}, 089903(E) (1998).

\bibitem{szekeres:obs2}
A. Nwankwo, M. Ishak and J. Thompson: Luminosity distance and redshift in the Szekeres inhomogeneous cosmological models.
\emph{J.C.A.P.} \textbf{05}, 028 (2011).

\bibitem{curvatureLTB} A. Paranjape and T.P. Singh: 
The possibility of cosmic acceleration via spatial averaging in Lema\^\i re--Tolman--Bondi models.
\emph{Class. Quant. Grav.} \textbf{23}, 6955 (2006). 

\bibitem{rasanen:peakmodel} S. R\"as\"anen: Evaluating backreaction with
the peak model of structure formation. \emph{J.C.A.P.} \textbf{04}, 026 (2008).

\bibitem{rasanen:model} S. R{\"a}s{\"a}nen: Cosmological acceleration from structure formation.
\emph{International Journal of Modern Physics D} \textbf{15}, 2141 (2006). 

\bibitem{rasanenFOCUS} S. R\"as\"anen: Backreaction: directions of progress. 
\emph{Class. Quant. Grav.} \textbf{28}, 164008 (2011).

\bibitem{rasanen:lightpropagation} S. R\"as\"anen:  
Light propagation and the average expansion rate in near--FRW universes. 
\emph{Phys. Rev. D} \textbf{85}, 083528 (2012).

\bibitem{cornelius}
C. Rampf and G. Rigopoulos: Zel'dovich approximation and general relativity.
\emph{Mon. Not. Roy. Astron. Soc.} \textbf{430}, L54 (2013).

\bibitem{boudDE}
B.F. Roukema, J.J. Ostrowski and T. Buchert: Virialization--induced curvature as a physical explanation for dark energy.
\emph{arXiv:1303.4444} (2013).

\bibitem{phasespace} 
X. Roy, T. Buchert, S. Carloni and N. Obadia: Global gravitational instability of FLRW backgrounds --- Interpreting  the dark sectors.
\emph{Class. Quant. Grav.} \textbf{28}, 165004 (2011).

\bibitem{roy:perturbations}
X. Roy and T. Buchert: Relativistic cosmological perturbation scheme on a general background: scalar perturbations for irrotational dust.
\emph{Class. Quant. Grav.} \textbf{29}, 115004 (2012).

\bibitem{russ:rza} 
H. Russ, M. Morita, M. Kasai and G. B{\"o}rner: Zel'dovich--type approximation for an inhomogeneous universe in general relativity: second--order solutions.
\emph{Phys. Rev. D}  \textbf{53}, 6881 (1996). 

\bibitem{russ:age} 
H. Russ, M.H. Soffel, M. Kasai and G. B{\"o}rner: Age of the universe: influence of the inhomogeneities on the global expansion factor. 
\emph{Phys. Rev. D} \textbf{56}, 2044 (1997). 

\bibitem{salopek}
D.S. Salopek, J.M. Stewart and K.M. Croudace: The Zel'dovich Approximation and the relativistic Hamilton--Jacobi Equation,
\emph{Mon. Not. Roy. Astron. Soc.} \textbf{271}, 1005 (1994).

\bibitem{silbergleit}
A.S. Silbergleit: Nonlinear motions against the Newtonian uniform expansion background: the case of the unperturbed density.
\emph{J. Math. Phys.} \textbf{36}, 847 (1995).

\bibitem{sussman} R.A. Sussman: Radial asymptotics of Lema\^\i tre--Tolman--Bondi dust models. 
\emph{Gen. Rel. Grav.} \textbf{42}, 2813 (2010).

\bibitem{sussman:review} R.A. Sussman: Back--reaction and effective acceleration in generic LTB dust models.
\emph{Class. Quant. Grav.} \textbf{28}, 235002 (2011).

\bibitem{matarrese:plane}
E. Villa, S. Matarrese and D. Maino: Post--Newtonian cosmological dynamics of plane--parallel perturbations and back--reaction.
\emph{J.C.A.P.} \textbf{08}, 024 (2011).

\bibitem{multiscale} A. Wiegand and T. Buchert: Multiscale cosmology and structure--emerging Dark Energy: a plausibility analysis. 
\emph{Phys. Rev. D.} \textbf{82}, 023523 (2010).

\bibitem{variance} A. Wiegand and D.J. Schwarz: Inhomogeneity--induced variance of cosmological parameters.
\emph{Astron. Astrophys.} \textbf{538}, A147 (2012).

\bibitem{wiltshireFOCUS}
D.L. Wiltshire: What is dust? -- Physical foundations of the averaging problem in cosmology.
\emph{Class. Quant. Grav.} \textbf{28}, 164006 (2011).

\end{thebibliography}
\end{document}